\documentclass[a4paper,fleqn,10pt]{article}
\usepackage{amsmath}
\usepackage{amssymb}
\usepackage[a4paper,pdfborder={0 0 0}]{hyperref}
\usepackage[all]{hypcap}
\usepackage{array}
\usepackage{calc}
\usepackage{longtable}
\usepackage{multirow}
\usepackage{axodraw}
\usepackage{pstricks}
\usepackage{graphicx}
\usepackage{xspace}
\usepackage{cite}
\usepackage{mcite}
\usepackage[format=hang,labelfont=bf]{caption}
\usepackage{sectsty}
\allsectionsfont{\sffamily}
\subsubsectionfont{\mdseries\itshape\large}
\let\spreprint\empty
\newcommand{\preprint}[1]{\def\spreprint{\protect#1}}
\let\sinstitute\empty
\newcommand{\institute}[1]{\def\sinstitute{\protect#1}}
\makeatletter
\renewcommand{\maketitle}{\begingroup
  \null\thispagestyle{empty}%
    \ifx\spreprint\empty
      \vskip 5ex
    \else
      \flushright\large\spreprint\vskip 2ex
    \fi
    \vskip 5ex
    \flushleft
      {\sffamily\bfseries\huge\@title}\vskip 2ex
      \@author\vskip 2ex
      \ifx\sinstitute\empty
      \else
        {\small\sinstitute}
      \fi
    \vskip 5ex
  \endgroup
}
\makeatother
\renewenvironment{abstract}{\begin{center}
  {\large\sffamily\bfseries Abstract: }
  \begin{minipage}[t]{0.75\textwidth}
}{\end{minipage}\end{center}\vskip 10ex}
\newenvironment{stress}{\vskip 2ex
  \hspace*{2ex}\begin{minipage}{\textwidth-4ex}
  \em}{\end{minipage}\vskip 2ex}
\newcommand{\myfigure}[3]{
  \begin{figure}[#1]
    \begin{center}
      #2\\
      \parbox[t]{\widthof{#2}}{\caption{#3}}
    \end{center}
  \end{figure}
}
\newcommand{\mytable}[3]{
  \begin{table}[#1]
    \begin{center}
      #2\\
      \parbox[t]{\widthof{#2}}{\caption{#3}}
    \end{center}
  \end{table}
}
\newcommand{\MCatNLO}{M\protect\scalebox{0.8}{C}@N\protect\scalebox{0.8}{LO}\xspace}
\newcommand{\HERWIG}{H\protect\scalebox{0.8}{ERWIG}\xspace}
\newcommand{\HERWIGpp}{H\protect\scalebox{0.8}{ERWIG++}\xspace}
\newcommand{\Ariadne}{A\protect\scalebox{0.8}{RIADNE}\xspace}
\newcommand{\POWHEG}{P\protect\scalebox{0.8}{OWHEG}\xspace}
\newcommand{\Sherpa}{S\protect\scalebox{0.8}{HERPA}\xspace}
\newcommand{\Comix}{C\protect\scalebox{0.8}{OMIX}\xspace}

\newcommand{\Amegic}{A\protect\scalebox{0.8}{MEGIC++}\xspace}

\newcommand{\CSS}{C\protect\scalebox{0.8}{SS}\xspace}
\newcommand{\Ahadic}{A\protect\scalebox{0.8}{HADIC++}\xspace}
\newcommand{\Hadrons}{H\protect\scalebox{0.8}{ADRONS++}\xspace}
\newcommand{\Photons}{P\protect\scalebox{0.8}{HOTONS++}\xspace}
\newcommand{\Pythia}{P\protect\scalebox{0.8}{YTHIA}\xspace}

\newcommand{\Professor}{P\protect\scalebox{0.8}{ROFESSOR}\xspace}
\newcommand{\Rivet}{R\protect\scalebox{0.8}{IVET}\xspace}
\long\def\symbolfootnote[#1]#2{\begingroup%
\def\thefootnote{\fnsymbol{footnote}}\footnote[#1]{#2}\endgroup}
\newcommand{\abs}[1]{\left| #1\right|}
\newcommand{\rbr}[1]{\left( #1\right)}
\newcommand{\abr}[1]{\langle #1\rangle}
\newcommand{\cbr}[1]{\left\{ #1\right\}}
\newcommand{\sbr}[1]{\left[ #1\right]}
\newcommand{\done}{{\rm d}}

\newcommand{\mc}[1]{\mathcal{#1}}

\newcommand{\dst}{\displaystyle}

\newcommand{\qcut}{Q_{\mathrm{cut}}}
\newcommand{\Nmax}{N_{\mathrm{max}}}
\newcommand{\GeV}{\mathrm{GeV}}

\newcommand{\kperpbfsq}{{\bf{k}}_{\perp}^{\,\!2}}

\newcommand{\logyc}{\log_{10} (\qcut^{2}/s)}

\hypersetup{
  pdfauthor={Stefan Hoeche, Frank Krauss,
    Steffen Schumann, Frank Siegert},
  pdftitle={QCD matrix elements and truncated showers},
  pdfkeywords={QCD, Matrix Element, Parton Shower,
    Truncated Shower, CKKW}
}
\preprint{ZU-TH 02/09\\IPPP/09/14\\DCPT/09/28\\
  HD-THEP-09-2\\MCNET/09/04}
\author{Stefan H{\"o}che$^1$, Frank Krauss$^2$,
  Steffen Schumann$^3$, Frank Siegert$^2$}
\title{QCD matrix elements and truncated showers}
\institute{$^1$ Institut f{\"u}r Theoretische Physik, 
  Universit{\"a}t Z{\"u}rich, CH-8057 Z{\"u}rich, Switzerland\\
  $^2$ Institute for Particle Physics Phenomenology,
  Durham University, Durham DH1 3LE, UK\\
  $^3$ Institut f{\"u}r Theoretische Physik, 
  Universit{\"a}t Heidelberg, D-69120, Heidelberg, Germany\\}
\begin{document}
\maketitle
\begin{abstract}
  We derive an improved prescription for the merging of matrix elements
with parton showers, extending the CKKW approach. A flavour-dependent 
phase space separation criterion is proposed. We show that this new 
method preserves the logarithmic accuracy of the shower, and that 
the original proposal can be derived from it. One of the main requirements 
for the method is a truncated shower algorithm. We outline the 
corresponding Monte Carlo procedures and apply the new prescription 
to QCD jet production in $e^+e^-$ collisions and Drell-Yan lepton pair
production. Explicit colour information from matrix elements
obtained through colour sampling is incorporated in the merging and
the influence of different prescriptions to assign colours in the 
large $N_C$ limit is studied. We assess the systematic
uncertainties of the new method.

\end{abstract}
\section{Introduction}
\label{sec:intro}
With the LHC becoming fully operational in the near future, searches
for new physics beyond the Standard Model (SM) will enter a new stage.  
Despite all optimism, a majority of the signals currently discussed 
suffers from severe backgrounds, among them many related to the 
production of heavy SM particles, such as the weak gauge bosons or 
top quarks, accompanied with jets.  Therefore, it is a central issue 
for many experimental analyses to correctly describe the production of these 
particles in conjunction with additional jets.  In many cases,
the method of choice is to employ simulation programs.  In the past, 
such event generators have proved to be extremely useful and versatile 
tools, being well capable to describe comparably simple event topologies 
at sufficiently high precision.  However, especially, when additional 
hard jets complicate the overall event structure, a proper simulation is 
far from being trivial.  Typically such extra radiation is approximated
with leading-logarithmic accuracy through the probabilistic description
provided by the parton-shower approach.  With rising precision needs, 
however, improved methods become mandatory, which describe the radiation 
of additional particles beyond the leading-logarithmic approximation 
employed by the parton showers.  

The most traditional of these methods consists in reweighting QCD 
emissions as described by the parton 
shower with respective exact matrix elements expressed through 
parton-shower variables~\cite{Seymour:1994df,
  *Corcella:1998rs,*Norrbin:2000uu}.  Unfortunately, 
the applicability of this method, however elegant, is constrained to 
cases, where the parton-shower expression either exceeds the matrix 
element or can be modified accordingly, without hampering the event generation
efficiency too badly.  This limits the method to a few cases (such as
the production of a gluon in $e^+e^-\to q\bar q$, top-quark decay plus
emission of an additional gluon, or the production of vector bosons in
hadron collisions).

In the past years, new and powerful methods for the systematic inclusion 
of higher order effects into event generation have been developed.  In 
fact, they can be seen as a major theoretical improvement in the detailed
understanding of complicated event topologies.  The first of these new
methods provides means to consistently match NLO calculations for
specific processes with the parton shower and has been incorporated into
the \MCatNLO program~\cite{Frixione:2002ik,*Frixione:2006gn}.  Its basic 
idea is to organise the counter-terms necessary to technically cancel real
and virtual infrared divergences in such a way that the first emission of 
the parton shower is recovered.  This allows the generation of hard 
kinematics configurations, which can eventually be fed into a parton-shower 
Monte Carlo. Several applications of the original approach to different 
processes have been presented, see for instance~\cite{Frixione:2003ei,*Frixione:2005vw,*Frixione:2008yi}.
A further improvement, aiming at an enhanced independence of both the 
specific hard process and the details of the parton-shower algorithm is 
provided by the \POWHEG-method~\cite{Nason:2004rx,*Frixione:2007vw}, 
which uses the ratio of the actual 
real radiation matrix element and the original leading-order one to generate 
the hardest emission.  In its current formulation, this approach can be 
understood as a hybrid of the traditional parton-shower reweighting and the 
\MCatNLO-method.  It has been implemented for various processes, see for 
example~\cite{Nason:2006hfa,*LatundeDada:2006gx,*Hamilton:2008pd,*LatundeDada:2008bv}.

An alternative approach, aiming at an improved description of multi-jet 
topologies, has been described in~\cite{Catani:2001cc,Krauss:2002up},
and is often called the CKKW algorithm.  It has been studied, e.g., 
in~\cite{Krauss:2004bs,Krauss:2005nu,*Gleisberg:2005qq} in the 
cases of $W$ and $Z$ production and the production of pairs of these 
bosons at the Tevatron and the LHC.  The idea here is to separate the 
phase space for parton emission into two domains, a hard region of jet 
production and a softer regime of intra-jet evolution.  This separation is 
achieved through a $k_\perp$-type jet measure~\cite{Brown:1991hx,*Catani:1991hj,
  Catani:1992zp,*Catani:1993hr}.  Then matrix elements 
for different parton multiplicities are used to describe the production 
of a corresponding number of jets, whereas the parton shower is constrained
such that it does not produce any additional jets.  Leading higher-order 
effects are added to the various matrix elements by reweighting them with 
appropriate Sudakov form factors and with ratios of the strong coupling 
$\alpha_s$, taken at the $k_\perp$-scales of the individual jet emissions.  
Independence of the overall result at logarithmic accuracy on the cut in 
the jet measure is achieved 
by the interplay of Sudakov form factors and the vetoed parton shower 
with suitable starting conditions. The method is one of the cornerstones of 
the event generator \Sherpa~\cite{Gleisberg:2003xi,*Gleisberg:2008ta}.
A similar approach, formally equivalent in $e^+e^-$ annihilations into 
hadrons and often called CKKW-L, has been formulated 
in~\cite{Lonnblad:2001iq,*Lavesson:2005xu}.  In this case a dipole shower
rather than a more traditional parton cascade is used to describe QCD 
radiation beyond fixed order.  Recently, first steps towards a merging at 
NLO precision of the matrix elements in this approach have been presented
in~\cite{Lavesson:2008ah}.  Another, more simplified method, often called 
the MLM-prescription, has been introduced in~\cite{Mangano:2001xp}. 
In \cite{Mangano:2006rw} the approach was applied for the case of top-quark 
production at hadron colliders. An extension of the algorithm to account for
hard QCD radiation in the production of heavy coloured states as they appear
in various beyond the Standard Model scenarios was presented in \cite{Alwall:2008qv}. 
The differences between the three merging prescriptions have been investigated using the 
example of $W$-production at the Tevatron and the 
LHC~\cite{Hoche:2006ph,*Alwall:2007fs}. Although the three methods are
different and the formal relationship of the latter method with the two former 
ones could not be completely worked out there, the results presented in 
these studies are in astonishing agreement.  First steps towards such a 
more formal comparison between the three algorithms has been attempted 
in~\cite{Lavesson:2007uu}. Recently, so-called ``multiplicative matching''
methods \cite{Bauer:2008qh,*Bauer:2008qj} have also attracted some attention.

In this paper we aim at establishing a general theory framework, in which 
different merging algorithms can be compared on formal grounds.  As a
result of this we also describe a merging algorithm that preserves the
logarithmic accuracy of the shower.  We prove this in the most general case,
including initial state QCD particles.  Although the prescriptions above deal 
with this case, up to now no statements of the formal accuracy beyond the
most leading logarithms has been made in the literature.  

Therefore, the outline of this paper is as follows: 
Section~\ref{sec:prerequisites} introduces the procedures necessary to 
consistently evolve parton showers from matrix elements of arbitrary
final-state multiplicity, which is one of the key ingredients of the envisaged 
merging approach.  Section~\ref{sec:merging} presents the general merging 
procedure.  It also sets the theoretical background of the new approach and 
introduces the event generation algorithm.  It is viable for, in principle, 
arbitrary parton-shower algorithms.  We show that the new procedure exactly 
reproduces the logarithmic accuracy of the shower.  The relationship of the
framework presented here with the other three algorithms named above is
discussed in Appendix~\ref{sec:ckkw}, where we show that the original CKKW 
approach as well as the CKKW-L algorithm can be derived as special cases.  
We also present results of our algorithm in two relevant cases, namely 
$e^+e^-$ annihilations into hadrons at LEP I and Drell-Yan-like production 
of gauge bosons at the Tevatron.  In Sec.~\ref{sec:megenerators} we describe 
the Monte Carlo programs, which are employed in this study.  
Its details, mainly concerning the systematics of the merging itself are
presented in Section~\ref{sec:comparison} and discussed in some detail. 
We conclude in Sec.~\ref{sec:conclusions}.

\section{Prerequisites for merging matrix elements and parton showers}
\label{sec:prerequisites}
Merging matrix elements with parton showers combines two essentially different approaches 
to perturbative QCD.  Hard matrix elements are exact at some fixed perturbative order 
in the strong coupling $\alpha_s$ and are therefore efficient in describing exclusive events
with fixed jet multiplicity, taking into account non-trivial interferences between 
different amplitudes.  Parton showers are employed to generate the QCD radiation pattern,
especially at lower scales, close to the hadronisation scale $\Lambda_{\rm QCD}$. 
Their application resums potentially large logarithmic corrections due to 
Bremsstrahlung effects. In their description they naturally cannot take into account 
all interferences, although steps towards including more quantum mechanical 
effects are currently being discussed, cf.~\cite{Nagy:2007ty,*Nagy:2008ns,
  *Nagy:2008eq}. In a simulation, best results can be expected, if the two approaches 
are combined consistently, such that each of them operates in those regions of phase 
space that it describes best. 

This necessitates that parton showers can evolve from parton configurations which are given
by high multiplicity matrix elements at certain points in phase space and colour space.
In such cases the starting conditions for parton showers are often ambiguous. We will see in
Sec.~\ref{sec:merging} that for running the shower in merged samples, it is necessary 
to interpret the matrix element as a core process and a series of predefined shower 
branchings. The reason is that the parton-shower evolution can take place at any point in this 
branching history, giving rise to the truncated shower prescription described in 
Sec.~\ref{sec:truncated_showers}. There may be various such histories arising from
the same matrix-element configuration. Hence the most appropriate one needs to be identified.
Typically, in the spirit of the probabilistic picture underlying the parton shower, it 
is simply assumed that this is the most probable one.

In this section we derive the formalism and the algorithms necessary to identify
the most probable shower history. We also explain the concept of truncated showers 
and elaborate on various strategies to define colour assignments for the shower input.

\subsection{Master evolution equations}
To prepare for the following arguments we firstly introduce
prototypical evolution equations for parton showers~\cite{Ellis:1991qj}, 
cf.~\cite{Collins:1981uk,*Collins:1981uw,*Collins:1987pm,*Collins:1989gx}
and~\cite{Konishi:1978yx,*Konishi:1979cb,*Kalinowski:1980ju}
\footnote{ The functions $g$ can also explicitly depend on the splitting variable
  $\zeta$, like in the case of angular ordered DGLAP evolution~\cite{Catani:1989ne,
  *Mueller:1981ex,*Ermolaev:1981cm,*Bassetto:1982ma,*Dokshitzer:1982xr}. This does,
  however, not complicate our formalism because corresponding terms do not modify the
  form of the Sudakov form factor. As far as such an evolution is concerned, the corresponding
  notation in the resolved term of Eq.~\eqref{eq:shower_evolution} is implicit.}
\begin{equation}\label{eq:shower_evolution}
  \frac{\partial g_a(z,t)}{\partial\log(t/\mu^2)}\,=
  \int_z^{\,\zeta_{\rm max}}\frac{\done\zeta}{\zeta}\,
  \sum\limits_{b=q,g}\mc{K}_{ba}(\zeta,t)\,g_b(z/\zeta,t)
  -g_a(z,t)\,\int_{\xi_{\rm min}}^{\,\xi_{\rm max}}
  \done\xi\,\sum\limits_{b=q,g}\xi\,\mc{K}_{ab}(\xi,t)\;.
\end{equation}
In this context, $g$ may denote either a fragmentation function (FF) or a parton 
distribution function (PDF), cf.\ for example~\cite{Amati:1978wx,*Amati:1978by}. 
The first term on the right hand side encodes resolvable emissions, 
while the second (related through unitarity) describes unresolved branchings 
and virtual contributions. The variable $t$ is the evolution parameter, 
while $\zeta$ is the splitting variable of the scheme.
The evolution kernels $\mc{K}_{ab}$ are obtained from ratios of appropriate $N+1$- and 
$N$-particle matrix elements in the respective limit. Schematically
\begin{equation}\label{eq:kernels_meratio}
  \mc{K}_{ab}(z,t)\to\frac{1}{\sigma_a^{(N)}(\Phi_N)}\,
    \frac{\done^2\sigma_b^{(N+1)}(z,t;\Phi_N)}{\done\log(t/\mu^2)\,\done z}\;.
\end{equation}
Here $\Phi_N$ denotes the respective $N$-particle phase space configuration,
which does not play a role for the limiting behaviour of $\sigma_b^{(N+1)}(z,t;\Phi_N)$.
Equation~\eqref{eq:kernels_meratio} conversely implies that any splitting kernel $\mc{K}_{ab}$ 
can be substituted by an appropriate ratio of matrix elements, because respective
differences are always subleading.
For the most common case of standard DGLAP evolution~\cite{Gribov:1972ri,*Lipatov:1974qm,
  *Dokshitzer:1977sg,*Altarelli:1977zs}, the kernels are easily identified through
\begin{equation}\label{eq:kernels_dglap}
  \mc{K}_{ab}(z,t)\to\frac{\alpha_s(z,t)}{2\pi}\,P_{ab}(z)\;,
\end{equation}
with $P_{ab}(z)$ being the standard DGLAP splitting functions. If 
Eq.~\eqref{eq:shower_evolution} is written in inclusive form, i.e.\ $\xi_{\rm min}\to 0$, 
$\xi_{\rm max},\zeta_{\rm max}\to 1$, the last term vanishes because of momentum 
sum rules for the kernels. In exclusive form, where the $\zeta$- and $\xi$-boundaries 
are determined by a resolution criterion for parton emission, it can be written as the 
logarithmic derivative of the Sudakov form factor~\cite{Sudakov:1954sw,*Korchemsky:1988hd}
\begin{equation}\label{eq:sudakov}
  \Delta_a(\mu^2,t)\,=\;\exp\cbr{\,-\int_{\mu^2}^{\,t}\frac{\done\bar t}{\bar t}
    \int_{\xi_{\rm min}}^{\,\xi_{\rm max}}\done\xi\sum\limits_{b=q,g}
    \frac{1}{2}\,\mc{K}_{ab}(\xi,\bar t)\,}\;.
\end{equation}
The factor $1/2$ is equivalent to $\xi$ under the integral and the sum over
parton species and avoids double-counting identical decay channels. 
Equation~\eqref{eq:sudakov} has the generic form of a no-emission 
probability between the scales $\mu^2$ and $t$. Potential differences
in various shower algorithms implementing QCD evolution arise due to different 
evolution kernels $\mc{K}$ or a different interpretation of the evolution and splitting 
variables. This in turn corresponds to the choice of a factorisation scheme.
If Eq.~\eqref{eq:shower_evolution} is modified accordingly,
the kernels can also incorporate more than two partons, for example in the case 
of dipole \cite{Gustafson:1986db,*Gustafson:1987rq,Lonnblad:1992tz,Giele:2007di,Winter:2007ye} 
or dipole-like \cite{Nagy:2005aa,*Nagy:2006kb,Dinsdale:2007mf,Schumann:2007mg} cascades.
Therefore, in this paper the term ``parton shower'' also includes the dipole or dipole-like
shower algorithms.

\subsection{Branching probabilities}
Equation~\eqref{eq:shower_evolution} is commonly rewritten in terms of $g/\Delta$.
The corresponding form reads
\begin{equation}\label{eq:shower_evolution_sud}
  \frac{\partial}{\partial\log(t/\mu^2)}\,
  \frac{g_a(z,t)}{\Delta_a(\mu^2,t)}=
  \frac{1}{\Delta_a(\mu^2,t)}
  \int_z^{\,\zeta_{\rm max}}\frac{\done\zeta}{\zeta}\,
  \sum\limits_{b=q,g}\mc{K}_{ba}(\zeta,t)\,g_b(z/\zeta,t)\;.
\end{equation}
This immediately yields the no-branching probabilities for unconstrained (forward) 
and constrained (backward) shower evolution as~\cite{Ellis:1991qj}
\begin{equation}\label{eq:sp_forward_general}
  \mc{P}^{(F)}_{{\rm no},\,a}(t,t')=
  \frac{\Delta_a(\mu^2,t')}{\Delta_a(\mu^2,t)}=
  \exp\cbr{-\int_t^{\,t'}\frac{\done \bar t}{\bar t}
    \int_{\zeta_{\rm min}}^{\,\zeta_{\rm max}}\done\zeta\,
      \sum_{b=q,g}\frac{1}{2}\,\mc{K}_{ab}(\zeta,\bar t)}\;,
\end{equation}
and
\begin{equation}\label{eq:sp_backward_general}
  \mc{P}^{(B)}_{{\rm no},\,a}(z,t,t')=
  \frac{\Delta_a(\mu^2,t')\,g_a(z,t)}{
    \Delta_a(\mu^2,t)\,g_a(z,t')}=
    \exp\cbr{-\int_t^{\,t'}\frac{\done \bar t}{\bar t} 
      \int_z^{\,\zeta_{\rm max}}\frac{\done\zeta}{\zeta} 
      \sum_{b=q,g}\mc{K}_{ba}(\zeta,\bar t)\,
      \frac{g_b(z/\zeta,\bar t)}{g_a(z,\bar t)}}\;.
\end{equation}
The standard procedure for constructing a parton-shower algorithm is to write
the differential branching probability $\mc{P}_{\rm branch,\,a}$ as 
\begin{equation}
\mc{P}_{\rm branch,\,a}(t,t')=\frac{\partial\mc{P}_{\rm no,\,a}(t,t')}{\partial\log(t/\mu^2)}\;.
\end{equation}
Given the current evolution scale $t'$, at which an actual emission took place, a new scale 
$t$ for the next parton emission is chosen according to this probability.

\subsection{Shower histories from matrix elements} 
\label{sec:cluster_algo}
An obvious point in the recursive algorithm sketched above is to define the first 
or hardest scale, at which the parton shower starts off.  In order to obtain such 
suitable starting conditions for parton showers from arbitrary
matrix elements, a clustering algorithm needs to be defined, which corresponds 
to ``running the shower evolution backwards'' on the respective matrix element. 
It identifies how, in a parton-shower picture, the matrix element would have 
originated from a lower multiplicity matrix element and a shower branching.
Applied iteratively, it leads to the definition of a core process, which cannot 
be further decomposed and a sequence of shower branchings yielding the
actual final state. The tasks for the algorithm are thus twofold: Firstly,
within an arbitrary $n$-parton final state the most probable splitting 
in terms of shower evolution starting with $n-1$ partons needs to be found. 
Secondly, corresponding partons must be recombined to obtain the respective 
final state. In order to construct this algorithm, the shower evolution simply 
has to be ``inverted'', which gives the following recipe:
\begin{stress}
The criterion of the cluster algorithm is defined by the 
shower evolution kernels.\\
The recombination scheme is given by the inverted shower kinematics.
\end{stress}
A side effect of this prescription is that during backward clustering the hard matrix 
elements, potentially no strict hierarchy is found in the shower evolution parameter. 
However, such situations can only arise for kinematic configurations of the matrix element,
which are beyond the accuracy of the shower.  Therefore they do not pose a problem here.

\subsection{Truncated showering}
\label{sec:truncated_showers}
Assume a parton-shower history constructed from a matrix element 
along the lines of Sec.~\ref{sec:cluster_algo} and consisting of a
core process plus a single additional branching at scale $t$, which we call
matrix-element branching. As a consequence of 
Eqs.~\eqref{eq:sp_forward_general} and~\eqref{eq:sp_backward_general}, parton
shower emissions may take place at scales $t'>t$. This leads to a situation, 
where, due to additional partons originating from these branchings, 
the kinematics of the matrix-element branching at $t$ needs to be redefined.
This corresponds to a truncated shower, identical to the situation discussed in
\cite{Nason:2004rx,*Frixione:2007vw}, where a mismatch between hardest emission, in terms
of transverse momentum, and the parton-shower evolution defined in opening
angles has been noted and resolved.  There the solution to this situation has been
coined ``truncated showering'', because the evolution stops at the lower, dynamical 
scale $t$, unlike ordinary shower evolution, which stops at the universal
cutoff scale. A prescription to unambiguously reconstruct the kinematics
of matrix-element branchings is needed. The most natural choice is to compute
the evolution, splitting and angular variables of matrix element branchings
in the shower scheme and reconstruct the branching using the shower kinematics,
after the final state of the branching at $t'>t$ is fully determined.
In other words:
\begin{stress}
Splitting kernels introduced by Eq.~\eqref{eq:kernels_meratio} define
evolution, splitting and angular variables in the shower scheme, rather than
the kinematics of the corresponding branching.
\end{stress}
This coincides with the fact that such branchings are interpreted
as predetermined nodes during shower evolution. It must simply be guaranteed
that the evolution stops at the corresponding scale, inserts the node
and restarts afterwards. If for any reason (e.g.\ energy-momentum conservation)
the matrix element branching cannot be reconstructed after a truncated shower
branching, this shower branching must be vetoed.

\subsection{Colour treatment}
\label{sec:colour}
The treatment of colour is a central issue when dealing with matrix element 
and shower generation in QCD processes. Shower formulations inherently are correct only 
at leading order in $1/N_C$, although in~\cite{Nagy:2007ty,*Nagy:2008ns,
  *Nagy:2008eq} first attempts to overcome this limitation have been presented. 
Therefore, once matrix elements are to be combined with showers, the treatment 
of colour must be adjusted.  A simple and obvious way to do so is to interpret the 
hard matrix element in the large-$N_C$ limit to define colour partners of splitting 
partons in the shower language. This problem is more easily solved, when colours are not
summed over but if an algorithm is employed, which unambiguously assigns a certain set 
of colours to the external particles in the hard matrix element. The basic idea is
then to sample over colours in a Monte Carlo fashion rather than summing over them.

It was pointed out in~\cite{Maltoni:2002mq,Duhr:2006iq} that the colour-flow 
decomposition of QCD amplitudes is advantageous over both the fundamental
and (if existent) the adjoint representation decomposition. The key point is 
that in the colour-flow decomposition each colour octet is treated as a $3\times3$ 
index field whose additional degree of freedom is removed by a projector onto
the colour-octet subspace. Since this issue is central to all our simulations
incorporating fixed colour assignments in the hard matrix element, we briefly 
recall the basics of the algorithm. As an example we consider an $n$-gluon
amplitude $\mc{A}\rbr{1,\ldots,n}$. This amplitude can be decomposed in the 
colour-flow basis as~\cite{Maltoni:2002mq}
\begin{equation}\label{eq:colour_decomposition_colourflow}
  \mc{A}\rbr{1,\ldots,n}=\sum\limits_{\sigma\in S_{n-1}}
    \delta^{i_1\bar{\jmath}_{\sigma_2}}
    \delta^{i_{\sigma_2}\bar{\jmath}_{\sigma_3}}
    \ldots\delta^{i_{\sigma_n}\bar{\jmath}_1}\;
    A\rbr{1,\sigma_2,\ldots,\sigma_n}\;.
\end{equation}
Here $i_{\sigma_k}$ and $\bar{\jmath}_{\sigma_k}$ denote the $3$- and 
$\bar{3}$-index of parton $\sigma_k$, respectively and the sum runs
over all possible permutations of the set $\cbr{2,\ldots,n}$.
The quantities $A\rbr{1,\sigma_2,\ldots,\sigma_n}$ are called colour-ordered
or partial amplitudes. They depend on the kinematics of the process only. 
All information about colour is incorporated in respective prefactors. 
Therefore any colour-ordered amplitude only contains planar diagrams, 
which greatly alleviates the computation. A convenient way to interpret 
Eq.~\eqref{eq:colour_decomposition_colourflow} is to consider it as the decomposition
of the full QCD amplitude into subamplitudes in the large-$N_C$ limit.
Feeding the results from the matrix-element calculation into a 
shower program, the corresponding colour connections are thus readily 
determined if one of the terms in the sum is picked as the most probable 
colour structure and identify the colour flow according to its colour factor.
In this context we use the fact that interference terms between
two different colour structures are always subleading in $1/N_C$.\footnote{
  This argument holds in the colour-flow decomposition and the 
  fundamental representation decomposition. For the latter, 
  see for example~\cite{Mangano:1987xk}.}

An algorithm to identify the most probable colour structure could thus look 
as follows (cf.~\cite{Caravaglios:1998yr})
\begin{enumerate}
\item Compute the full matrix element with randomly assigned colours
  for external QCD partons.
\item\label{csalgo_step2}
  Identify all possible permutations $\cbr{1,\vec{\sigma}}$, which 
  yield a non-zero value of
  \begin{equation}
    \delta^{i_1 \bar\jmath_{\sigma_2}}
    \delta^{i_{\sigma_2} \bar\jmath_{\sigma_3}}
    \cdots\delta^{i_{\sigma_n} \bar\jmath_1}\;.
  \end{equation}
  Label them by $\vec{\sigma}_i$ and
  compute the corresponding partial amplitudes $A(1,\vec{\sigma}_i)$.
\item If $N_{\vec{\sigma}}$ is the number of identified permutations,
  choose a partial amplitude with probability
  \begin{equation}
    P_{\vec{\sigma}_i}=
    \frac{\abs{A(1,\vec{\sigma}_i)}^2}
         {\sum_{j=1}^{N_{\vec{\sigma}}}\abs{A(1,\vec{\sigma}_j)}^2}
  \end{equation}
\end{enumerate}
Because of the way potential partial amplitudes are identified in the
colour-flow decomposition, this prescription is similar to the following
simplified strategy
\begin{enumerate}
\item Compute the full matrix element with randomly assigned colours
  for external QCD partons.
\item
  Assign colours in the large-$N_C$ limit at random, respecting
  the actual point in colour space. This translates into two partons only being
  colour adjacent at large $N_C$, if they were colour adjacent at
  finite $N_C$. 
\item 
  Identify the corresponding permutation $\vec{\sigma}$ and compute 
  the partial amplitude $A(1,\vec{\sigma})$.
  Accept the configuration with probability
  \begin{equation}
    P_{\vec{\sigma}}=\frac{\abs{A(1,\vec{\sigma})}^2}{
      \abs{\mc{A}(1,\ldots,n)}^2}
  \end{equation}
\end{enumerate}
Naively, the drawback of the latter algorithm is, that potentially many 
points have to be drawn for the colour assignment at large $N_C$. 
In practice, however, this is sufficiently fast compared to evaluating all 
possible partial amplitudes for a single nontrivial point in colour space. 
Also, in principle the full amplitude squared,
$\abs{\mc{A}(1,\ldots,n)}$, might be much smaller than the sum of
partial amplitudes squared, such that acceptance probabilities are
modified. This algorithm is still sufficiently accurate, since respective 
differences are always subleading in $1/N_C$.

The above algorithm works for arbitrary QCD and QCD-associated matrix 
elements, since quark pairs can always be reinterpreted as colour octet objects.
Matrix-element configurations might exist, which do not allow an immediate
projection onto large $N_C$ because of the U(1) pseudo-gluon. In this
case, a new point in colour space can safely be assigned, because the
respective contribution to the total cross section is subleading.

\section{The merging algorithm}
\label{sec:merging}
The central idea for algorithms merging matrix elements with parton showers is 
to replace products of splitting kernels related to hard emissions in the shower 
evolution with the appropriate matrix elements, thus reinstalling information 
about the full hard process under consideration. Directly implementing a ratio of 
hard matrix elements in form of a splitting kernel has the apparent disadvantage 
that the respective phase-space integration proceeds in terms of shower kinematics 
and is thus hard to optimise in a generic way. A better technique is to first compute 
the matrix element and then reweight it such that, to the accuracy of the parton shower, 
the corresponding shower expression is obtained. To pursue this strategy, 
the corresponding no-emission probabilities of the parton shower, i.e.\ its Sudakov form 
factors, must be known. This can, however easily be achieved because they emerge directly
from the evolution equations on which the shower is based. 

Only one additional ingredient is eventually needed, namely a criterion, which defines 
how to separate matrix-element and parton-shower domain. It will be shown in 
Sec.~\ref{sec:jet_criterion}
that a general form of this criterion can be found, which is based on the soft and 
collinear behaviour of QCD at the next-to-leading order. We will refer to it
as the ``jet criterion''. At the present stage, the precise form of the jet criterion is 
unimportant and it is sufficient and helpful to think of it in an abstract way.

\subsection{Construction of the algorithm}
The basic idea of the merging -- to separate the matrix-element and parton-shower 
domains through a cut in the emission phase space, defined by a jet criterion -- 
corresponds to a simple phase-space slicing. We therefore define the evolution kernels
for the matrix-element and parton-shower domains
\begin{align}\label{eq:kernel_meps}
  \mc{K}^{\rm ME}_{ab}(\xi,\bar t)=&\;\mc{K}_{ab}(\xi,\bar t)\;
    \Theta\sbr{\vphantom{\sum}Q_{ab}(\xi,\bar t)-Q_{\rm cut}}
  &&\text{and}
  &\mc{K}^{\rm PS}_{ab}(\xi,\bar t)=&\;\mc{K}_{ab}(\xi,\bar t)\;
    \Theta\sbr{\vphantom{\sum}Q_{\rm cut}-Q_{ab}(\xi,\bar t)}\;,
\end{align}
where $Q_{ab}$ denotes the jet criterion and $Q_{\rm cut}$ is its cut value. 
Correspondingly, the two partial Sudakov form factors are given by
\begin{equation}\label{eq:sudakov_meps}
  \Delta^{\rm ME/PS}_a(\mu^2,t)=\exp\cbr{-\int_{\mu^2}^{\,t}
  \frac{\done\bar t}{\bar t}\int_{\xi_{\rm min}}^{\,\xi_{\rm max}}\done\xi\,
  \sum_{b=q,g}\frac{1}{2}\,\mc{K}^{\rm ME/PS}_{ab}(\xi,\bar t)\,}\;.\\
\end{equation}
They are related to the full Sudakov form factor, Eq.~\eqref{eq:sudakov}, through
\begin{equation}\label{eq:sudakov_factorisation}
  \Delta_a(\mu^2,t)=\Delta^{\rm PS}_a(\mu^2,t)\,\Delta^{\rm ME}_a(\mu^2,t)\;.
\end{equation}
In fact, Eq.~\eqref{eq:sudakov_factorisation} effectively encodes the complete 
merging approach. There, ultimately $\mc{K}^{\rm ME}$ will be replaced with a ratio
of matrix elements, according to Eq.~\eqref{eq:kernels_meratio}. During the following 
rewrite of the evolution equations we simply identify, how the factorisation property 
of Sudakov form factors must then be interpreted and employed for event generation. 
No further proof for the correctness of the algorithm at any logarithmic order is 
necessary, because this follows directly from the accuracy implemented 
in the parton-shower evolution. In other words, the proposed merging scheme does not 
impair the logarithmic accuracy of the parton shower.

This can be summarised as:
\begin{stress}
When correcting parton-shower evolution kernels through higher-order matrix elements, 
the master evolution equation, Eq.~\eqref{eq:shower_evolution}, must be respected. 
This ensures that the accuracy generated by the factorisation scheme and the parton 
shower is fully restored.
\end{stress}

We start by defining the conditional backward no-branching probability
in the parton-shower domain\footnote{From here on we focus on backward evolution.
  The corresponding reasoning for forward evolution follows trivially.},
\begin{equation}\label{eq:sp_backward_ps}
  \tilde{\mc{P}}^{(B)\,{\rm PS}}_{{\rm no},\,a}(z,t,t')=
  \frac{\Delta^{\rm PS}_a(\mu^2,t')\,\tilde{g}_a(z,t)}{
    \Delta^{\rm PS}_a(\mu^2,t)\,\tilde{g}_a(z,t')}=
    \exp\cbr{-\int_t^{\,t'}\frac{\done \bar t}{\bar t} 
      \int_z^{\,\zeta_{\rm max}}\frac{\done\zeta}{\zeta} 
      \sum_{b=q,g}\mc{K}^{\rm PS}_{ba}(\zeta,\bar t)\,
      \frac{\tilde{g}_b(z/\zeta,\bar t)}{\tilde{g}_a(z,\bar t)}}\;.
\end{equation}
It corresponds to the modified evolution equation
\begin{equation}\label{eq:ps_evolution_sud}
  \frac{\partial}{\partial\log(t/\mu^2)}\log
  \frac{\tilde{g}_a(z,t)}{\Delta^{\rm PS}_a(\mu^2,t)}=
  \int_z^{\,\zeta_{\rm max}}\frac{\done\zeta}{\zeta}\,
  \sum\limits_{b=q,g}\mc{K}^{\rm PS}_{ba}(\zeta,t)\,
  \frac{\tilde{g}_b(z/\zeta,t)}{\tilde{g}_a(z,t)}\;.
\end{equation}
Equation~\eqref{eq:sp_backward_ps} differs with respect to the 
standard parton-shower evolution because of the $\Theta$-function, restricting 
emissions to $Q<Q_{\rm cut}$, i.e.\ the soft and collinear domain.
Its interpretation is therefore straightforward and gives a rule 
for the modified shower algorithm in the merging:
\begin{stress}
Standard parton-shower evolution is implemented, 
but radiation with $Q>Q_{\rm cut}$ is vetoed.
\end{stress}
Note that for the case of an initial-state parton shower, typically described with
a backward evolution, the initial scale of the PDF's is set by the core process of 
the event.

If Eq.~\eqref{eq:sp_backward_ps} is employed as is, 
including the phase-space restriction, the newly defined 
functions $\tilde{g}$ do not obey the same evolution as
the original functions $g$. Factorisation is thus violated. 
If we want the two evolutions to agree, we have to guarantee that the full 
no-branching probability in the merging approach is given by
Eq.~\eqref{eq:sp_backward_general}. This leads to the definition 
of the no-emission probability in the matrix-element domain according to
\begin{equation}\label{eq:me_nobranch_sud}
  \mc{P}^{(B)\,\rm ME}_{{\rm no},\,a}(t,t')=
  \frac{\mc{P}^{(B)}_{{\rm no},\,a}(z,t,t')}{
    \mc{P}^{(B)\,\rm PS}_{{\rm no},\,a}(z,t,t')}=
  \frac{\Delta^{\rm ME}(\mu^2,t')}{\Delta^{\rm ME}(\mu^2,t)}\;,
  \quad\text{where}\quad
  \mc{P}^{(B)\,{\rm PS}}_{{\rm no},\,a}(z,t,t')=
  \frac{\Delta^{\rm PS}_a(\mu^2,t')\,g_a(z,t)}{
    \Delta^{\rm PS}_a(\mu^2,t)\,g_a(z,t')}\;.
\end{equation}
It is interesting to note that $\mc{P}^{(B)\,\rm ME}_{{\rm no},\,a}$ is independent 
of $z$, which effectively is an outcome of the factorisation 
properties of PDF's and FF's.
  
If we assume that a ``most probable'' shower history of the hard 
matrix element is identified through a backward-clustering algorithm, 
which employs the clustering criterion and the recombination scheme of the shower, 
cf.\ Sec.~\ref{sec:cluster_algo}, then we obtain the rule:
\begin{stress}
The weight, Eq.~\eqref{eq:me_nobranch_sud}, is assigned to 
any leg with production scale $t'$ and decay scale $t$ found during 
backward clustering. Strong couplings are evaluated
at the nodal scales of parton recombination.
\end{stress}
The reasoning is easily explained. Hard matrix elements
in the factorisation scheme of the shower have the same limiting
behaviour as the splitting kernels $\mc{K}$, once colour-adjacent partons
become close in phase space. Backward clustering 
will identify a hierarchical structure for the factorisation of
hard matrix elements into lower-multiplicity matrix elements
and splitting kernels. Eventually, a core process is found, 
which cannot be further decomposed and which corresponds to the 
starting conditions for a respective shower evolution. 
Matrix elements, however, do not implement the no-branching 
probabilities generated by parton showers. Also the strong coupling
is evaluated at a common scale, rather than the nodal scales of
splittings. Corresponding corrections must therefore be implemented.

An immediate consequence of the above defined algorithm
is that the total cross section of merged samples can only 
be influenced by the difference between full hard matrix elements
and the corresponding product of splitting kernels times the core 
process. In this respect, we obtain the rule:
\begin{stress}
To compute hadronic cross sections, PDF's must be evaluated
at the scale of the core process defined through backward clustering.
\end{stress}
This prescription is independent of the multiplicity of the
matrix element, because showering always starts at the scale 
of the core process, cf.\ Eq.~\eqref{eq:sp_backward_general}. 
A mismatch in the two scales would lead to 
ill-defined backward no-branching probabilities.

\subsection{Event generation techniques}
\label{sec:algorithm}
Event generation according to the above defined merging procedure
proceeds along the following lines:
\begin{enumerate}
\item[$\bullet$] Relevant multi-jet cross sections for the process under consideration
  are calculated with the phase-space restriction $Q>Q_{\rm cut}$. Strong couplings 
  are computed such that they give an overestimate, which can later on be reweighted. 
  PDF's are evaluated at the scale set by the core process.
\item\label{algostep1} Events are generated according to the above defined 
  cross sections with kinematics determined by the respective matrix elements.
\item Matrix elements are interpreted in the large $N_C$ limit 
  according to methods discussed in Sec.~\ref{sec:colour}.
  The most probable shower history of the final state is determined
  through backward clustering, cf.\ Sec.~\ref{sec:cluster_algo}. 
  The clustering is guided by information 
  from the matrix element, which restricts the available shower histories 
  to those, which correspond to a Feynman diagram.
\item\label{algostep3}
  The event is accepted or rejected according to a kinematics-dependent 
  weight, which corresponds to evaluating strong couplings in the shower scheme
  and computing the no-branching probability, Eq.~\eqref{eq:me_nobranch_sud},
  for each parton, internal or external, in the clustered matrix element.
\item\label{algostep5} The parton-shower evolution is started with suitably defined
  scales for intermediate and final-state particles. During showering, any emission
  harder than $Q_{\rm cut}$ is vetoed. Intermediate partons undergo a
  truncated shower evolution, cf.\ Sec.~\ref{sec:truncated_showers}
\end{enumerate}

This algorithm has the apparent drawback, that the no-emission probability
Eq.~\eqref{eq:me_nobranch_sud} must be computed before the parton-shower evolution 
starts. Ideally, however, it should result as a direct consequence of parton-shower 
branchings producing hard partons. Such splittings would obviously fall in the realm of matrix
elements and are thus forbidden inside the parton shower. To obtain a corresponding 
prescription, the above algorithm is slightly reformulated.

Firstly, the logarithmic derivative of the no-branching probabilities
$\mc{P}^{(B)\,\rm ME/PS}_{{\rm no},\,a}$ is defined as
\begin{align}\label{eq:ik_meps_common_evolution}
  \mc{I}^{(B)\,\rm ME/PS}_a(z,\bar t)=&\;\int_z^{\,\zeta_{\rm max}}
  \frac{\done\zeta}{\zeta}\sum_{b=q,g}\,\mc{K}^{\rm ME/PS}_{ba}(\zeta,\bar t)\,
  \frac{g_b(z/\zeta,\bar t)}{g_a(z,\bar t)}\;.
\end{align}
From Eq.~\eqref{eq:sp_backward_general} one then obtains the full branching
probability in terms of $\mc{I}^{(B)\,\rm ME/PS}$,
\begin{equation}\label{eq:bp_meps_common_evolution}
  \mc{P}^{(B)\,\rm ME\oplus PS}_{{\rm branch},\,a}(z,t,t')=
  \sbr{\;\mc{I}^{(B)\,\rm ME}_a(z,t)+\mc{I}^{(B)\,\rm PS}_a(z,t)\;}\,
  \exp\cbr{-\int_t^{\,t'}\frac{\done\bar t}{\bar t}\;
    \mc{I}^{(B)}_a(z,\bar t)\,}\;,
\end{equation}
where $\mc{I}^{(B)}_a=\mc{I}^{(B)\,\rm ME}_a+\mc{I}^{(B)\,\rm PS}_a$.
Equation~\eqref{eq:bp_meps_common_evolution} corresponds to generating an 
ordering parameter $t$ in unconstrained shower evolution, i.e.\ without the
restriction to $Q<Q_{\rm cut}$. The first term in the square bracket is however 
given by hard matrix elements through Eq.~\eqref{eq:kernels_meratio}. 
In order not to double count this contribution, corresponding branchings 
must lead to rejection of the entire event.
This modifies the respective cross section by 
\begin{equation}\label{eq:xs_reweighting_onestep_algo}
  \sigma\to\sigma\,\cdot\,\mc{P}^{(B)\,\rm ME}_{{\rm no},\,a}(t,t')\;.
\end{equation}
Due to this event rejection, the remaining branching probability for accepted 
parton-shower steps is given by (cf.\ the description of the veto algorithm,
for example in~\cite{Sjostrand:1995iq})
\begin{equation}\label{eq:vetoed_ps_onestep_algo}
  \begin{split}
  \mc{P}^{(B)\,\rm PS}_{{\rm branch},\,a}(z,t,t')=&\;
  \mc{I}^{(B)\,\rm PS}_a(z,t)\,
  \exp\cbr{-\int_t^{\,t'}\frac{\done\bar t}{\bar t}\;
    \mc{I}^{(B)}_a(z,\bar t)\,}\\
  &\qquad\times\exp\cbr{\;\int_t^{\,t'}\frac{\done\bar t}{\bar t}\;
    \sbr{\,\mc{I}^{(B)}_a(z,\bar t)-\mc{I}^{(B)\rm PS}_a(z,\bar t)\,}\,}\;,
  \end{split}
\end{equation}
which yields the vetoed shower algorithm described by
Eq.~\eqref{eq:sp_backward_ps}, but with $\tilde{g}=g$, as required.\\
We therefore obtain the modified rules
\begin{enumerate}
\item[3.]\label{algomstep3}
  The event is accepted or rejected according to a kinematics dependent 
  weight, which corresponds to evaluating strong couplings in the parton-shower scheme.
\item[4.]\label{algomstep5} The parton-shower evolution is started with suitably defined
  scales for intermediate and final-state particles. Intermediate partons undergo 
  truncated evolution. During showering, any emission harder than $Q_{\rm cut}$ leads to 
  the rejection of the event. 
\end{enumerate}
Note that in prinicple these two steps could be combined through evaluating the strong 
couplings during the shower evolution.

The relation of this algorithm with other merging prescriptions is discussed in
Appendix~\ref{sec:relation}.

\subsection{Highest multiplicity treatment}
\label{sec:highest_multi}
An apparent problem of the merging algorithm outlined so far is that only a
limited final-state multiplicity $N\le N_{\rm max}$ can be generated 
through full matrix elements. Hence the matrix elements will not produce 
jet multiplicities $N>N_{\rm max}$ that are in principle possible.  Hence, the
parton shower must account for missing emissions above $Q_{\rm cut}$ at large $N$. 
This is explained in detail in the following.

Assume a case where $N=N_{\rm max}$ emissions in the matrix-element
domain have been accounted for by the matrix element and have been
generated through the above defined algorithm.
This means that up to this point, i.e.\ up to the scale where the last 
matrix-element emission can be resolved in terms of the parton-shower 
evolution parameter, the branching probability,
Eq.~\eqref{eq:bp_meps_common_evolution} has been employed, as it should be.
Beyond this point, no further emission can be generated through matrix 
elements, and the branching probability becomes
\begin{equation}\label{eq:bp_meps_common_hm_faulty}
  \mc{P}^{(B)\,\rm ME\oplus PS}_{{\rm branch},\,a}(z,t,t')\to\;
  \mc{P}^{(B)\,\rm ME}_{{\rm no},\,a}(t,t')\,
  \frac{\partial\mc{P}^{(B)\,\rm PS}_{{\rm no},\,a}(z,t,t')}{\partial\log(t/\mu^2)}\;.
\end{equation}
Relation~\eqref{eq:bp_meps_common_hm_faulty} would obviously violate
factorisation, because of missing terms, corresponding to the 
integrated kernel from the matrix-element domain.

This problem can be circumvented by implementing the standard parton-shower
evolution beyond the last matrix-element emission.\footnote{
  The term ``beyond'' refers to the ordering parameter $t$.
  Note that the respective scale is set globally for the event, because
  the matrix element connects all parton-shower evolutions.
} 
It guarantees that the parton shower respects the description of hard radiation 
throughout the regime where matrix elements are applicable, while still 
filling the remaining phase space.

This prescription is referred to as the highest multiplicity treatment
and has been suggested in a similar form in~\cite{Krauss:2004bs}.
In virtuality ordered DGLAP evolution, it
approximately corresponds to setting a local veto scale 
$Q_{\rm cut}\to Q_{\rm min}$ if $N=N_{\rm max}$, where 
$Q_{\rm min}$ is the minimum jet criterion found during backward 
clustering.

\subsection{Sources of uncertainties}
\label{sec:uncertainties}
The proposed merging algorithm combines two essentially different approaches 
to perturbative QCD. Any simulation based on it therefore contains 
a number of sources of theoretical uncertainties. They can be separated into
two categories, merging-related and non-merging-related uncertainties.
The latter would occur in standard perturbative approaches as well, 
when using only hard matrix elements or applying only parton showers.
The merging-related uncertainties are instead specific for the combined
application of matrix elements and showers and arise from the following:
\begin{itemize}
\item The specific choice of the jet criterion.\\
  Since the jet criterion separates matrix-element and parton-shower domain,
  a variation of its precise definition can enhance or reduce the
  contribution of the hard matrix elements in certain regions of phase space.
\item The value of the phase-space separation cut, $\qcut$.\\
  As for the jet criterion itself, the precise value of the separation cut
  enhances or reduces the amount of phase space which is described by hard
  matrix elements and can therefore lead to variation of the results.
\item The maximum number of jets simulated by hard matrix elements, $\Nmax$.\\
  This parameter limits the number of hard partons up to which correlations
  can be expected to be correctly described at tree-level.
\end{itemize}
Other uncertainties are related to the perturbative calculations
carried out in the matrix-element and parton-shower simulation itself.
They include:
\begin{itemize}
\item Scale uncertainties from matrix elements.\\
  They arise due to the particular choice of factorisation
  and renormalisation scale of the leading-order process.
\item Scale uncertainties from parton showers.\\
  They arise due to the particular choice of coupling
  scales within the evolution.
\item Uncertainties from the parton density functions employed.\\
  Parton density functions not only enter the cross section calculation when 
  considering hadronic initial states but also appear in the calculation of the 
  branching probabilities for the initial-state parton shower, for a detailed 
  discussion of the latter see e.g.~\cite{Gieseke:2004tc}. 
\item Uncertainties due to the choice of the leading-order process.\\
  These uncertainties arise in processes which potentially contain 
  many additional jets with shower evolution parameters above 
  the factorisation scale of the leading-order process.
  Corresponding details are discussed in 
  Appendix~\ref{sec:multicore_merging}. 
\end{itemize}
In this publication we focus on a study of the pure
merging-related uncertainties. We will, however not vary the
jet criterion, but rather employ what we identify as the optimal choice 
for the merging, see Sec.~\ref{sec:jet_criterion}.

\section{The jet criterion}
\label{sec:jet_criterion}
An important aspect in the QCD evolution equations, Eq.~\eqref{eq:shower_evolution},
is that QCD branchings are logarithmically enhanced at small values of the
evolution parameter $t$ and/or at logarithmically large values of the evolution 
kernels $\mc{K}$. This is the manifestation of the singular infrared
behaviour of QCD amplitudes in the respective regions of phase space.
In perturbative calculations employing hard matrix elements these regions therefore 
must be regularised.  This is typically achieved by identifying parton samples or 
individual partons with jets and demanding the jets to be sufficiently isolated. 
Algorithms defining jets are, for example, the Durham 
$k_T$-algorithm~\cite{Brown:1991hx,*Catani:1991hj}
and the longitudinally invariant $k_T$ algorithms for deep inelastic scattering
and hadron-hadron collisions~\cite{Catani:1992zp,*Catani:1993hr}.
Extensions of those algorithms to include jet flavour have been presented in~\cite{Banfi:2006hf}.
Their respective measures are often used as a variable in which
phase-space separation is defined for matrix element - parton shower merging,
cf.~\cite{Catani:2001cc,Krauss:2002up}.

We propose a similar criterion here. However, in contrast to jet measures like the 
ones above, which can be applied to experimentally observable final states 
and which yield experimentally well defined jets,
this criterion is designed and applied on purely theoretical grounds.
It is based on the behaviour of QCD amplitudes at the next-to-leading 
order and employs flavour and colour information of the respective partons.
For our purposes this criterion proved to be advantageous over standard 
$k_T$ algorithms as it correctly identifies individual infrared enhanced 
QCD branchings. 

\subsection{Definition}
Consider two partons $i$ and $j$, which can, in terms of flavour and colour, originate 
from a common mother parton (the splitter) $\widetilde{ij}$. The following jet criterion is 
then proposed
\begin{equation}\label{eq:jet_criterion}
  Q_{ij}^2\,=\;2\,p_i p_j\,\min\limits_{k\ne i,j}\,
    \frac{2}{C_{i,j}^k+C_{j,i}^k}\;,
\end{equation}
where for final state partons $i$ and $j$
\begin{equation}\label{eq:cij}
  C_{i,j}^k\,=\;\left\{\begin{array}{cc}
    {\dst \frac{p_ip_k}{(p_i+p_k)p_j}-\frac{m_i^2}{2\,p_ip_j}\,}
    &\text{if $j = g$}\\[1.5em]
    1 &\text{else}\\
    \end{array}\right.\;.
\end{equation}
For initial state partons $a$, we consider the splitting process
$a\to\rbr{aj}\,j$. With the momentum of the combined particle $\rbr{aj}$
given by $p_{aj}=p_a-p_j$, we define
\begin{equation}\label{eq:cij_is}
  C_{a,j}^k\,=\;C_{\rbr{aj},\,j}^k\;.
\end{equation}
The minimum in Eq.~\eqref{eq:jet_criterion} is over all possible colour partners 
$k$ of the combined parton $\widetilde{ij}$, which can be thought of to act
as spectators in the splitting process.

In the following, it is shown that this jet criterion indeed correctly identifies 
soft and collinear parton splittings in QCD matrix elements and is thus suited 
to separate the matrix-element from the parton-shower domain in the merging.

\subsection{Soft limit}
If the energy of a single gluon $j$ tends to zero in any fixed direction $q$, 
described through $p_j=\lambda q$, $\lambda\to 0$, the above jet criterion 
behaves as
\begin{equation}\label{eq:jet_criterion_soft}
  \frac{1}{Q_{ij}^2}\to\frac{1}{2\,\lambda^2}\;\frac{1}{2\,p_i\, q}\,
    \max_{k\ne i,j}\sbr{\frac{p_ip_k}{(p_i+p_k)\,q}-\frac{m_i^2}{2\,p_iq}\,}\;.
\end{equation}
The corresponding singularity of the matrix element is thus correctly 
identified, cf.~\cite{Catani:2002hc}.

\subsection{Quasi-collinear limit for final-state splittings}
Consider two final-state partons $i$ and $j$ and an arbitrary spectator-parton $k$.
Let $p_{ij}=p_i+p_j$ and let the light-like helper vectors $l$ and $n$ be defined by
\begin{equation}
  \begin{split}
  p_{ij}\,=&\;l+\alpha_{ij}\, n\;,\\
  p_k\,=&\;n+\alpha_k\, l\;.
  \end{split}
\end{equation}
This system has the solution
\begin{align}\label{eq:sudakov_ln}
  l\,=&\;\frac{1}{1-\alpha_{ij}\alpha_k}\,(p_{ij}-\alpha_{ij}\, p_k)\;,
  &n\,=&\;\frac{1}{1-\alpha_{ij}\alpha_k}\,(p_k-\alpha_k\, p_{ij})\;,
\end{align}
where $\alpha_{ij}=p_{ij}^2/\gamma$, $\alpha_k=p_k^2/\gamma$ and
$\gamma=2\,ln=p_{ij}p_k+\sqrt{(p_{ij}p_k)^2-p_{ij}^2 p_k^2}$, 
cf.~\cite{Pittau:1996ez,*Pittau:1997mv}.
The momenta $p_i$ and $p_j$ can now be expressed in terms of $l$, $n$ 
and a transverse component, $k_\perp$.
\begin{equation}\label{eq:jet_criterion_collinear_kinematics_fs}
  \begin{split}
  p_i^\mu\,=&\;z\,l^\mu+\frac{m_i^2+{\rm k}_\perp^2}{z}\,
    \frac{n^\mu}{2\,ln}+k_\perp^\mu\;,\\
  p_j^\mu\,=&\;(1-z)\,l^\mu+\frac{m_j^2+{\rm k}_\perp^2}{1-z}\,
    \frac{n^\mu}{2\,ln}-k_\perp^\mu\;.
  \end{split}
\end{equation}
A relation for $p_{ij}^2$ is immediately obtained,
\begin{equation}\label{eq:pipj_lcm}
  p_{ij}^{\,2}-m_i^2-m_j^2\,=\;\frac{{\rm k}_\perp^2}{z(1-z)}
    -\frac{1-z}{z}\,m_i^2-\frac{z}{1-z}\,m_j^2\;.
\end{equation}
Taking the quasi-collinear limit amounts to the simultaneous
rescaling~\cite{Catani:2000ef}
\begin{align}
  {\rm k}_\perp\to&\lambda{\rm k}_\perp\;,
  &m_i\to&\lambda m_i\;, 
  &m_j\to&\lambda m_j\;,
  &m_{ij}\to&\lambda m_{ij}\;.
\end{align}
Then, $2\,p_ip_j\to\lambda^2 (p_{ij}^{\,2}-m_i^2-m_j^2)$ and, independent of $k$,
\begin{equation}\label{eq:jet_criterion_collinear_fs}
  \frac{1}{Q_{ij}^2}\to\frac{1}{2\,\lambda^2}\;
  \frac{1}{p_{ij}^{\,2}-m_i^2-m_j^2}\,
    \rbr{\,\tilde{C}_{i,j}+\tilde{C}_{j,i}\,}\;.
\end{equation}
Here,
\begin{equation}\label{eq:cij_collinear}
  \tilde{C}_{i,j}=\left\{\begin{array}{cc}
      {\dst\frac{z}{1-z}-\frac{m_i^2}{2\,p_ip_j}}&\text{if $j=g$}\\[1.5em]
      1&\text{else}
    \end{array}\right.\;.
\end{equation}
Equation~\eqref{eq:cij_collinear} corresponds to the leading term of the massive 
Altarelli-Parisi splitting function for $z\to 1$~\cite{Catani:2000ef}. 
The corresponding term for $z\to 0$ (if present) is generated by $\tilde{C}_{j,i}$.

\subsection{Quasi-collinear limit for initial-state splittings}
Now consider the initial-state parton $a$, the final-state parton $j$ 
and an arbitrary spectator parton $k$. Let $p_{aj}=p_a-p_j$, and let
the light-like helper vectors $l$ and $n$ be defined by
\begin{equation}
  \begin{split}
  p_a\,=&\;l+\alpha_a\, n\;,\\
  p_k\,=&\;n+\alpha_k\, l\;.
  \end{split}
\end{equation}
Then $l$ and $n$ are found as before, Eq.~\eqref{eq:sudakov_ln}. The momenta
$p_{aj}$ and $p_j$ are decomposed as follows
\begin{equation}\label{eq:jet_criterion_collinear_kinematics_is}
  \begin{split}
  p_{aj}^\mu\,=&\;z\,l^\mu+\frac{p_{aj}^2+{\rm k}_\perp^2}{z}\,
    \frac{n^\mu}{2\,ln}+k_\perp^\mu\;,\\
  p_j^\mu\,=&\;\rbr{1-z}\,l^\mu
    +\frac{m_j^2+{\rm k}_\perp^2}{1-z}\,
    \frac{n^\mu}{2\,ln}-k_\perp^\mu\;.
  \end{split}
\end{equation}
Taking the quasi-collinear limit yields
$2\,p_ap_j\to\lambda^2 \abs{p_{aj}^{\,2}-m_a^2-m_j^2}$ such that, independent of $k$
\begin{equation}\label{eq:jet_criterion_collinear_is}
  \frac{1}{Q_{aj}^2}\to\frac{1}{2\,\lambda^2}\;
  \frac{1}{\abs{p_{aj}^{\,2}-m_a^2-m_j^2}}\,
    \rbr{\,\tilde{C}_{\rbr{aj},j}+\tilde{C}_{j,\rbr{aj}}\,}\;,
\end{equation}
where $\tilde{C}_{\rbr{aj},j}$ is given by Eq.~\eqref{eq:cij_collinear}.

\section{Monte Carlo programs for the study}
\label{sec:megenerators}
In this section, we present the Monte Carlo programs that have been employed
in the actual implementation of the merging algorithm outlined in
Sec.~\ref{sec:merging}. As discussed in Sec.~\ref{sec:uncertainties},
the foremost aim for this publication is to study the
specific systematics of the proposed merging procedure. 

For tree-level amplitudes there exists now a full wealth of
publicly available programs that accomplish the evaluation of 
(nearly) arbitrary multi-parton processes within the Standard Model 
and some of its most prominent extensions. These highly automated 
tools, called matrix-element or parton-level generators, either rely 
on the translation of Feynman diagrams into helicity 
amplitudes~\cite{Belanger:2003sd,*Alwall:2007st} or make use of recursive 
relations to obtain compact expressions for the 
amplitudes~\cite{Kanaki:2000ey,Mangano:2002ea,*Kilian:2007gr}. In addition to 
the pure matrix elements they provide suitable phase-space integrators to permit the 
evaluation of cross sections and the generation of actual parton
level events. In the context of this work we employ two different matrix-element
generators, the well-established program \Amegic~\cite{Krauss:2001iv}
and the recently released code \Comix~\cite{Gleisberg:2008fv}. Both codes are 
integrated into the multi-purpose event generator \Sherpa~\cite{Gleisberg:2003xi,*Gleisberg:2003ta}
that steers the event generation and hosts various interfaces to parton
showers, hadronisation routines, analysis methods and the like. The two 
programs share the capability to simulate arbitrary Standard Model 
processes at tree-level. Their main difference lies in the treatment 
of colour, providing us with a handle on the related systematic uncertainties.
We will briefly repeat the main features of both approaches in Secs~\ref{sec:megenerators:amegic} 
and Sec.~\ref{sec:megenerators:comix}, respectively.

Contrary to the matrix-element generators, there exist only few publicly available
implementations of parton-shower algorithms. There is the class of \Pythia-like 
parton showers using either virtuality \cite{Sjostrand:2003wg,*Kuhn:2000dk,*Krauss:2005re} or 
transverse momentum \cite{Sjostrand:2004ef} as evolution variable. On the 
other hand there are implementations of angular-ordered parton showers in the \HERWIG 
and \HERWIGpp generators, see for instance~\cite{Corcella:2002jc,*Bahr:2008pv} and 
references therein. A shower implementation based on colour dipoles became 
available with the release of the \Ariadne~\cite{Lonnblad:1992tz} program.
For our study we employ the parton-shower model presented in \cite{Schumann:2007mg}. This 
approach relies on Catani--Seymour dipole factorisation of QCD amplitudes and organises 
subsequent parton emissions in terms of an invariant transverse momentum. Aspects of 
this parton-shower model relevant for combining it with matrix element calculations 
will be discussed in Sec.~\ref{sec:showers}.

\subsection{The matrix-element generator \texorpdfstring{\Amegic}{AMEGIC++}}
\label{sec:megenerators:amegic}
\Amegic~\cite{Krauss:2001iv} is a tree-level matrix-element generator, based on Feynman
diagrams. To evaluate 
the single amplitudes of a given process, the helicity methods introduced 
in~\cite{Kleiss:1985yh,*Ballestrero:1994jn} are employed.
Diagrams are constructed and sorted according to their respective colour 
structure. A colour matrix for the full squared matrix element is computed 
using standard methods. Each single Feynman diagram is then decomposed into 
basic building blocks in the helicity formalism. It is important to note in this context, that 
information about the diagrams can be accessed during the runtime of \Sherpa 
and therefore clustering sequences within the merging formalism can be chosen 
according to the propagator structure of the contributing graphs. Thus,
unphysical combinations can be prevented. The fact that the treatment of colour 
in \Amegic essentially follows textbook methods is however rather
problematic in the context of a matrix element - parton shower merging.
Since it is not possible within the code to access colour-ordered amplitudes,
a ``most probable'' colour structure must be selected on kinematic grounds.
This then yields the nodal values to define starting conditions for parton showers.

\subsection{The matrix-element generator \texorpdfstring{\Comix}{COMIX}}
\label{sec:megenerators:comix}
\Comix is based on an extension of the colour-dressed Berends-Giele 
recursive relations for QCD amplitudes~\cite{Duhr:2006iq} to the full
Standard Model. These relations are essentially equivalent to the 
Dyson-Schwinger recursion employed for instance in~\cite{Kanaki:2000ey}. 
For any recursive calculation, the growth in 
computational complexity of the algorithm solely depends on the number 
of external legs at elementary vertices of the theory. Thus within 
\Comix{} any four-particle vertex of the Standard Model is decomposed into 
three-particle vertices~\cite{Gleisberg:2008fv}. This leads to an improved performance
for large-multiplicity matrix elements, compared to \Amegic. The summation (averaging) 
over colours in QCD and QCD-associated processes is performed 
in a Monte Carlo fashion and colour-ordered amplitudes can therefore 
be computed. Following the reasoning of~\cite{Duhr:2006iq}, 
the colour-flow basis is employed throughout the code. As discussed 
in~\cite{Maltoni:2002mq} and Sec.~\ref{sec:colour}, this yields a certain 
correspondence between the large-$N_C$ limit employed in parton-shower simulations 
and full QCD results. \Comix allows to access current information
during event generation, such that, like for \Amegic, unphysical clusterings
in the matrix element - parton shower merging can be prevented. The corresponding
algorithm respects the actual colour assignment of external states, 
cf.\ Sec.~\ref{sec:colour}.

\subsection{The Catani-Seymour subtraction based shower generator}
\label{sec:showers}
In this work we employ a shower approach based on Catani-Seymour (CS)
dipole factorisation, which will be denoted by \CSS~\cite{Schumann:2007mg}. 
The model was originally proposed in~\cite{Nagy:2005aa,*Nagy:2006kb} 
and worked out and implemented in~\cite{Dinsdale:2007mf,Schumann:2007mg}. 
It relies on the factorisation of real-emission matrix elements in the 
CS subtraction framework~\cite{Catani:1996vz,Catani:2002hc}. 
There exist four general types of CS dipole terms that capture the complete
infrared singularity structure of next-to-leading order QCD amplitudes, namely splitting
initial-state particles accompanied by an initial or final-state parton as spectator, 
or branching final-state lines associated with either another final-state leg or
an incoming parton as spectator. In the large-$N_C$ limit, the splitter and spectator 
partons are always adjacent in colour space. The dipole functions for the various 
cases, taken in four dimensions and averaged over spins, are used as shower 
splitting kernels. Their infrared singularities are regularised through a 
finite cutoff parameter of order $\Lambda_{\rm QCD}$, the shower stopping scale. 
Consider, as an example, the case of a splitting initial-initial CS dipole, 
cf.\ Fig.~\ref{fig:split_II}. Following the nomenclature used in~\cite{Schumann:2007mg}, 
we can then, in analogy to Eq.~\eqref{eq:kernels_dglap}, identify
\begin{equation}\label{eq:kernels_css}
\mc{K} \to \mc{K}_{a(ai)}(x_{i,ab},\kperpbfsq) = 
\frac{\alpha_s(\kperpbfsq/4)}{2\pi}\,\abr{{\bf V}^{ai,b}(x_{i,ab})}\;.
\end{equation}
\myfigure{t}{\hspace*{3.5cm}
  \begin{picture}(200,100)(0,0)
  \Line(100,50)( 10, 20)
  \Line(100,50)(175, 20)
  \LongArrow( 20,16)( 46, 25)
  \LongArrow( 54,46)( 34, 66)
  \LongArrow(170,30)(140, 42)
  \Line( 55,35)( 20,70)
  \Vertex( 55,35){2.5}
  \GCirc(100,50){10}{1}
  \put( 73, 25){$\widetilde{ai}$}
  \put(  5, 10){$a$}
  \put( 20, 75){$i$}
  \put(180, 10){$b$}
  \put( 83, 67){$\abr{{\bf V}^{ai,b}}$}
  \put(157, 43){$p_b$}
  \put( 32, 10){$p_a$}
  \put( 48, 60){$p_i$}
  \end{picture}\hspace*{3.5cm} }{Schematic view of the splitting of an initial-state 
        parton with an initial-state 
        spectator.  The blob denotes the $n$-parton matrix element. Incoming and 
        outgoing lines label initial- and final-state partons, respectively.
  \label{fig:split_II}}

The actual splitting is parametrised in terms of the Lorentz invariants
\begin{eqnarray}
x_{i,ab}=\frac{p_ap_b-p_ip_a-p_ip_b}{p_ap_b}\,\quad {\rm and} \quad \
\kperpbfsq = 2\tilde p_{ai}p_b\,\tilde v_i\,\frac{1-x_{i,ab}}{x_{i,ab}}\,, \quad {\rm where} \quad \tilde v_i = \frac{p_ip_a}{p_ap_b}\;.\label{eq:CS_sv_II}
\end{eqnarray}

The invariant transverse momentum $\kperpbfsq$ acts as ordering parameter for initial-state 
splittings in the actual shower algorithm\footnote{
  Note that the actual form of $\kperpbfsq$ is different to the one in~\cite{Schumann:2007mg}, 
  allowing to suppress an additional Jacobian factor obtained ibidem.  We employ this same 
  quantity as ordering parameter for both dipole types with initial-state splitters, whereas 
  the choice for dipoles with final-state splitter remains unchanged with respect 
  to~\cite{Schumann:2007mg}, namely the invariant transverse momentum squared of the two 
  splitting products.
}. The three other CS dipole configurations are treated following a similar reasoning as 
just presented for initial-initial dipoles. Corresponding Sudakov form factors for all 
branching types, taking into account finite parton masses and strictly relying on 
Lorentz-invariant variables, have been derived in~\cite{Schumann:2007mg}. 

All branchings in the \CSS formalism implement exact four-momentum conservation and the 
particles after an evolution step are kept on their mass-shell. The phase-space maps from an
$n$ to an $n+1$ particle final state are exact and cover the whole phase space. With the only 
exception of splitting CS initial-initial dipoles, the momentum recoil of a certain branching 
is compensated locally by the assigned spectator parton. For splitting initial-initial dipoles 
the spectator momentum is kept fixed, consequently the recoil is taken by the entire ensemble 
of final-state particles. 
 
The recoil strategy in this shower scheme, in conjunction with the strictly Lorentz-invariant 
evolution parameters and splitting variables, eventually allows the construction of a 
clustering algorithm along the lines of Sec.~\ref{sec:cluster_algo} and a truncated shower, 
cf.\ Sec.~\ref{sec:truncated_showers}.
Both are necessary ingredients of the merging formalism presented in this publication. 
The main difference with respect to other shower formulations, in this respect, 
lies in the fact that the recoil partner of a splitting parton is always 
a single external line of the current initial or final state\footnote{This is
  also true for initial-initial dipoles, since the shower formulation is invariant
  and the reference frame can be chosen as the centre-of-mass frame of the process
  before the splitting.}.
If this is not guaranteed, the implementation of a clustering algorithm as proposed in 
Sec.~\ref{sec:cluster_algo} becomes cumbersome.

Another apparent advantage of this parton-shower model is that it inherently respects 
QCD soft colour coherence. By construction in Catani--Seymour factorisation the eikonal
factor associated to soft gluon emission off a colour dipole, used to derive the angular
ordering constrained in conventional parton showers, is exactly mapped onto two CS dipoles. 
These two dipoles just differ by the role of emitter and spectator, see~\cite{Catani:1996vz}. 
Accordingly, the shower algorithm based on Catani--Seymour dipole factorisation captures
the essential contributions of the next-to-leading logarithmic corrections, beyond the naive 
resummation of the leading soft and collinear enhanced terms. For a further discussion on 
related issues, the reader is referred to~\cite{Dokshitzer:2008ia,*Nagy2009re}. 
For a comparison of the shower model at hand with experimental data sensitive to the correct 
modelling of soft-gluon emission see~\cite{Schumann:2007mg}.

A potential shortcoming of the shower algorithm based on Catani--Seymour dipole terms is 
that certain dipole functions connecting the initial and final state may assume negative
values in some non-singular phase-space regions. This prohibits their naive interpretation in 
terms of splitting probabilities, cf.~\cite{Dinsdale:2007mf,Schumann:2007mg}.
This problem, however, can now partially be cured once shower emissions are corrected by
exact matrix-element calculations that provide the appropriate description of QCD
amplitudes in these regions of phase space.

\section{Results}
\label{sec:comparison}
In this section we present results generated with the Monte Carlo
programs described previously.
The main purpose of this brief study is to quantify some systematic
uncertainties introduced by the proposed merging algorithm, 
cf.\ Sec.~\ref{sec:uncertainties}. Therefore,
three main questions are considered:
\begin{itemize}
\item Are observables sufficiently independent on the ``unphysical''
  phase-space separation criterion $\qcut$?
\item How does the description of additional jets by the matrix element improve
  the prediction of experimentally relevant quantities?
\item As a combination of the above:
  Is the total cross section as predicted by the lowest order
  unaltered to the required accuracy?
\end{itemize}
Using the two matrix-element generators \Amegic and \Comix, together with the
shower generator \CSS in the framework of \Sherpa, we also have the opportunity to study
effects of using different methods to assign colours in the large $N_C$ limit.

In the first part of the section we focus on $e^+e^-$ annihilation into hadrons.
This setup allows to study the algorithm in pure final-state QCD evolution. 
The influence of different colour-assignment strategies is investigated in this configuration.
The second part deals with Drell-Yan lepton pair production in hadronic interactions,
which is used to validate the algorithm in combined initial and final-state evolution.

Jets at parton level include the quark flavours $d$, $u$, $s$, $c$ 
and $b$ as well as gluons. The PDF set employed is CTEQ~6L~\cite{Pumplin:2002vw}, which
defines the corresponding $\alpha_s$ parametrisation in hadron collisions.
All other generator parameters are left at the default values of the Monte Carlo
programs, since none of them has any impact on the QCD predictions.

Hadron-level results are produced using \Sherpa with the fragmentation module
\Ahadic~\cite{Winter:2003tt}, the hadron and $\tau$ decay package \Hadrons~\cite{hadrons}
and a simulation of extra QED radiation through \Photons~\cite{Schoenherr:2008av}.
The hadronisation phases remain untuned so far and will be tuned to data at a later stage 
with the help of the \Professor package~\cite{Buckley:2009ad}. 
The \Rivet package~\cite{Waugh:2006ip,*Buckley:2008vh} is employed for analysis and 
comparison to data.

\subsection{QCD jet production in \texorpdfstring{$e^+e^-$}{e+e-} collisions}
\label{sec:results_ee}
To compare with LEP data from the Run~I period, a setup with cms energy $\sqrt{s}=91.25 \,\GeV$
is chosen and a merged sample of $e^+e^- \to (N+2)\, {\rm jets}$ is produced at parton level,
with $N$ denoting the number of additional jets in the matrix element and $N\leq \Nmax$.
We vary $\Nmax$ between zero, i.e.\ no merging at all, and four.

\subsubsection*{Total cross sections and jet rates}
Firstly, we present a comparison of total cross sections predicted by the merging algorithm
for various values of the separation criterion $\qcut$ and the maximum jet multiplicity $\Nmax$.
Table~\ref{tab:eexs} shows only minor deviations, i.e.\ up to 6.4\%, between the leading-order
cross section and predictions for the merged samples. We can thus confirm that the
proposed merging approach preserves the cross section of the leading-order process.
\newcommand{\mrt}[1]{\multirow{3}{*}{#1}}
\newcommand{\mco}[1]{\multicolumn{1}{|c|}{#1}}
\newcommand{\mct}[1]{\multicolumn{2}{c|}{#1}}
\mytable{tbp}{
\begin{tabular}{cc|c|r@{}l|r@{}l|r@{}l|r@{}l|}
\cline{3-11}
                     &       & \multicolumn{9}{c|}{$\Nmax$} \\ \cline{3-11}
                     &       & 0              & \mct{1}    & \mct{2}    & \mct{3}    & \mct{4}    \\ \hline
\mco{\mrt{$\logyc$}} & -1.25 & \mrt{40.17(1)} & 39&.65(3)  & 39&.66(3)  & 39&.66(3)  & 39&.67(3)  \\ \cline{2-2}\cline{4-11}
\mco{}               & -1.75 &                & 39&.38(5)  & 39&.29(6)  & 39&.13(5)  & 39&.13(5)  \\ \cline{2-2}\cline{4-11}
\mco{}               & -2.25 &                & 39&.27(8)  & 38&.35(9)  & 37&.89(11) & 37&.60(10) \\ \hline
\end{tabular}
}{Total cross sections for $e^+ e^- \to jj$ at $\sqrt{s}=91.25 \mathrm{GeV}$ in [nb]
  and their dependence on the separation criterion and the maximum number of
  additional jets produced in the matrix element.
  \label{tab:eexs}}

\begin{figure}
\begin{centering}
  \includegraphics[width=0.485\linewidth]{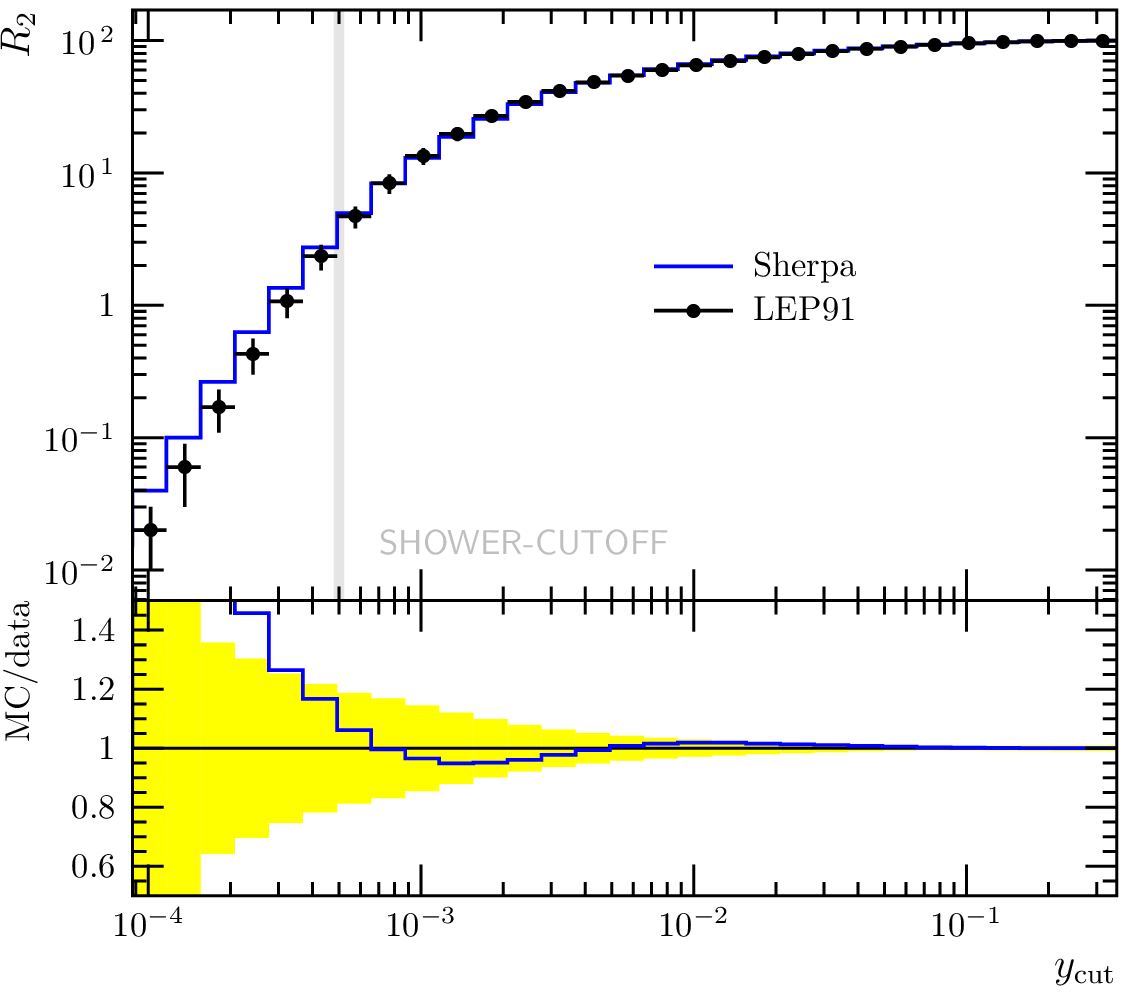}\hspace*{5mm}
  \includegraphics[width=0.485\linewidth]{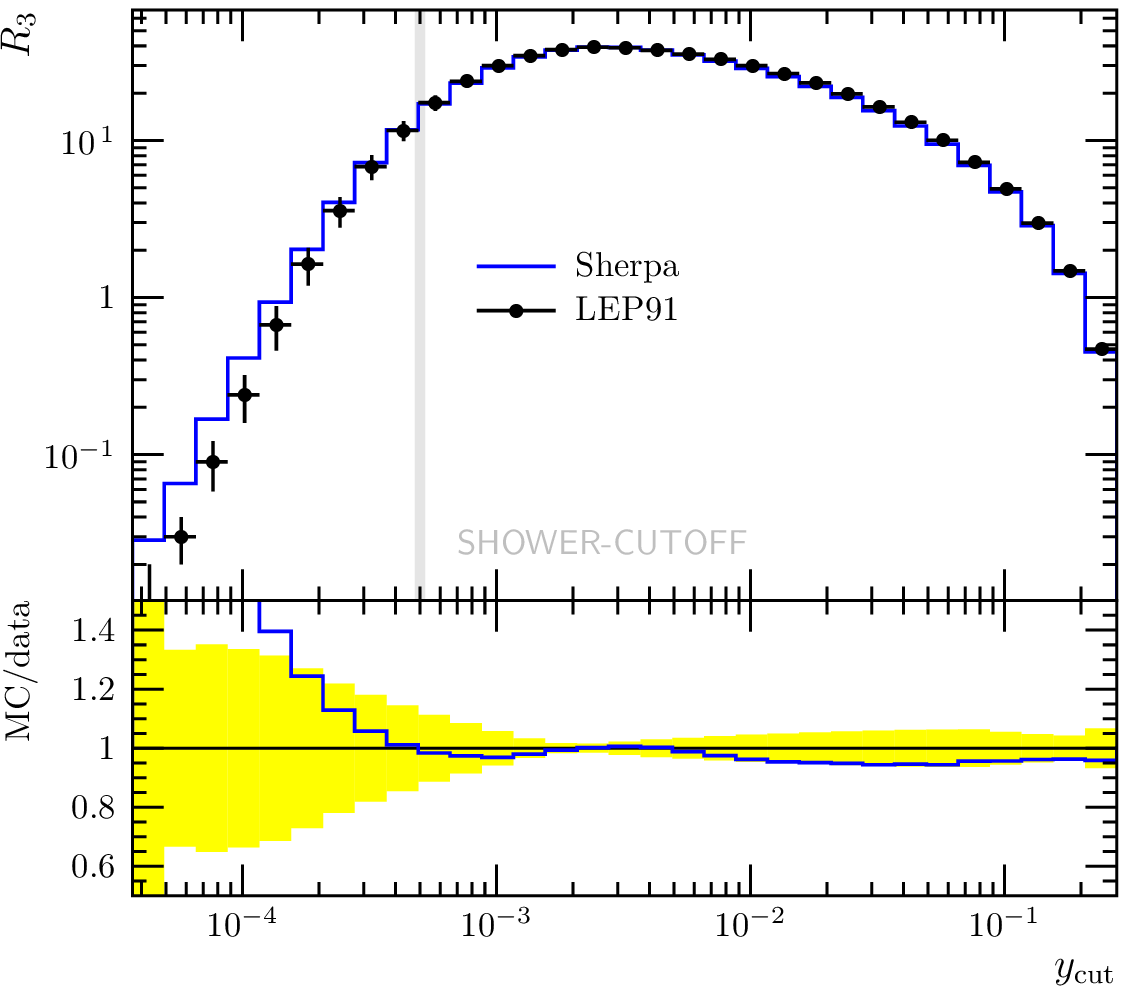}\vspace*{5mm}\\
  \includegraphics[width=0.485\linewidth]{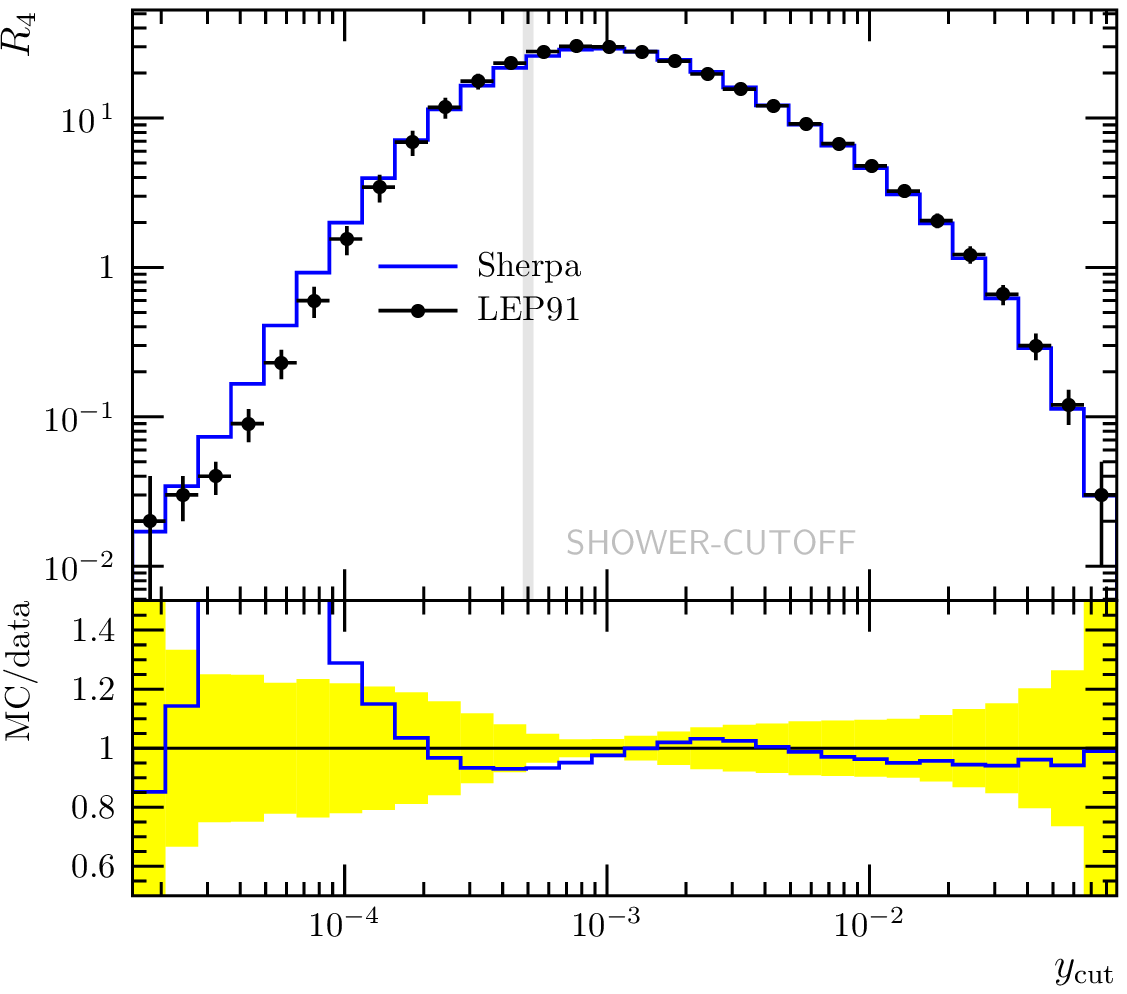}\hspace*{5mm}
  \includegraphics[width=0.485\linewidth]{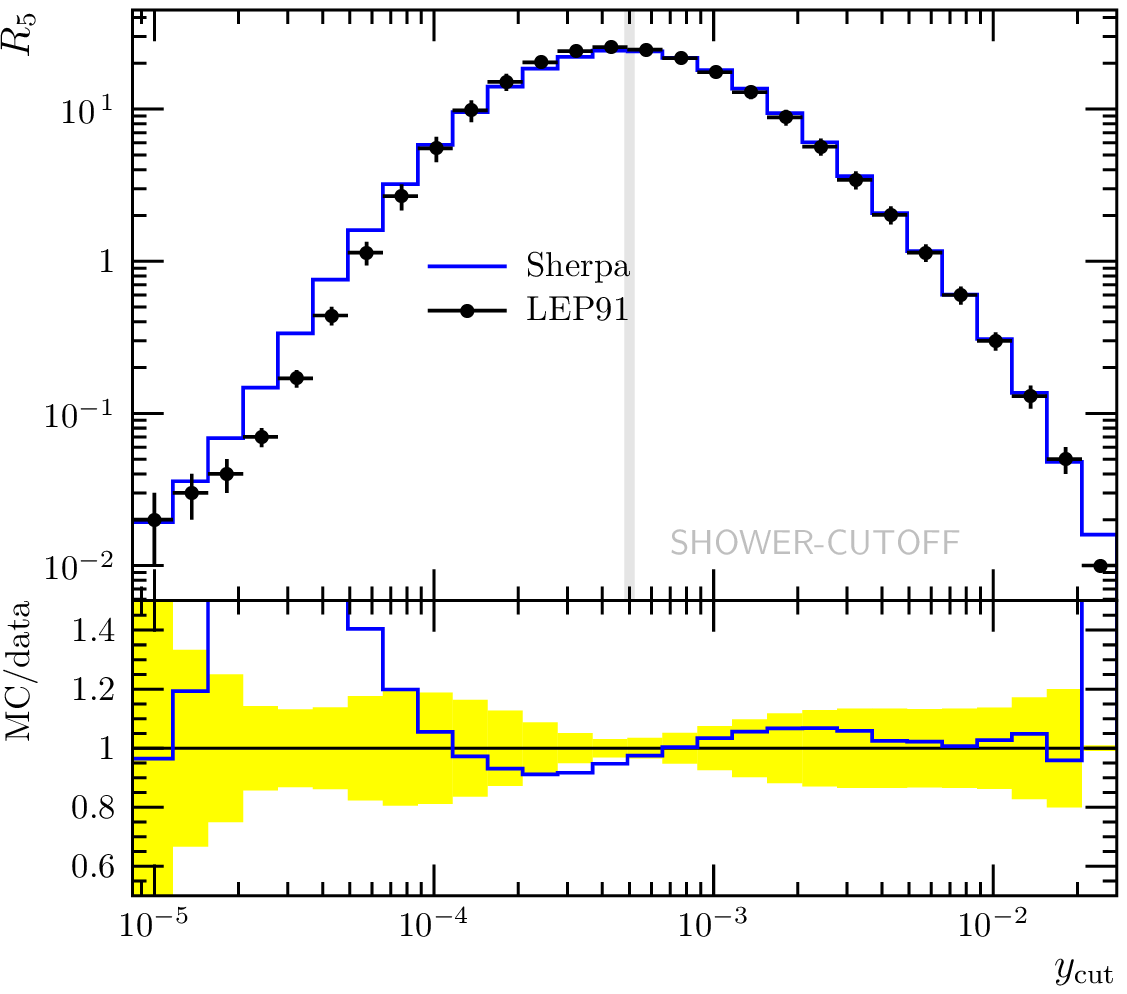}
  \caption{Integrated jet rates compared to data from LEP\protect\cite{Pfeifenschneider:1999rz}.}
  \label{fig:eenjetrates}
\end{centering}
\end{figure}

Figure~\ref{fig:eenjetrates} shows integrated rates of jets determined with the
Durham $k_T$-algorithm~\cite{Brown:1991hx,*Catani:1991hj} as a function of the
analysis cut $y_{\rm cut}$.
As well as giving a fine-grained insight into the number of jets to be
expected for a given analysis cut, they also provide the assurance that the
merging algorithm gives accurate predictions for the perturbative region in
which it operates. Jets found with cuts below the shower regime,
$y_{\rm cut} \approx 10^{-3.5}$, are influenced by hadronisation
effects, and are thus not relevant within the scope of this paper.
Even scales slightly above might be populated e.g. by decays of heavy mesons.
Monte Carlo results have been produced using a merged sample for up to four additional 
jets in the final state,
generated with \Comix and showered with \CSS with a merging cut $\logyc=2.25$.

\subsubsection*{Differential distributions}
Due to the close correspondence
between the jet criterion, Eq.~\eqref{eq:jet_criterion}, and the Durham
measure for jets in $e^+e^-$ collisions (cf.\ also Appendix~\ref{sec:ckkw}),
differential jet rates are particularly suited to assess merging systematics. 
The rate $y_{n\,n+1}$ shows the jet measure at which $n+1$ jets 
are clustered into $n$ jets according to the Durham $k_T$-algorithm.
The phase-space separation cut $\qcut$ manifests itself as a narrow 
transition region between matrix-element and parton-shower domain around 
$y_{n\,n+1} \approx \qcut^2/s$.
Merging systematics can thus be inferred from deviations between samples 
with different phase-space separation cuts in this region.

\begin{figure}
\begin{centering}
  \includegraphics[width=0.485\linewidth]{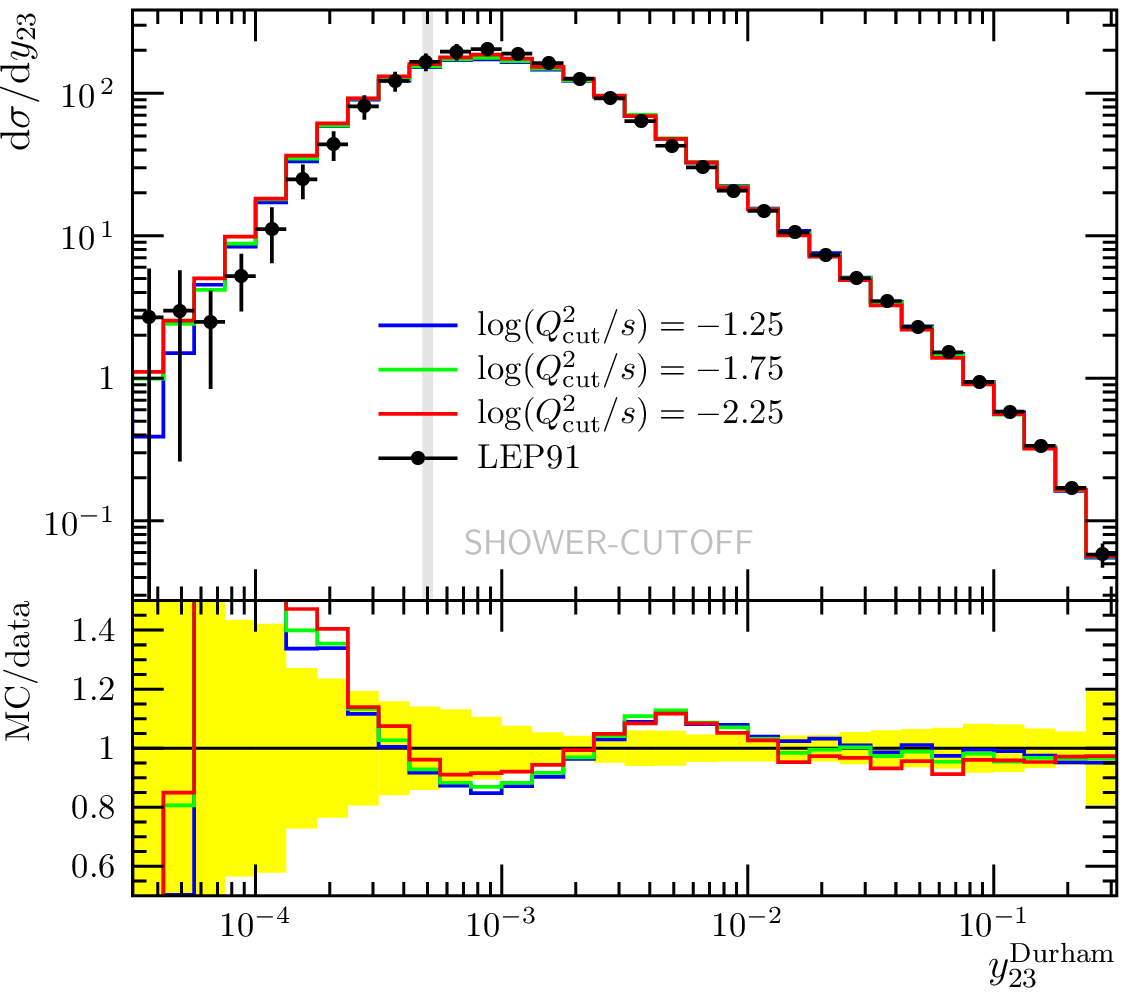}\hspace*{5mm}
  \includegraphics[width=0.5\linewidth]{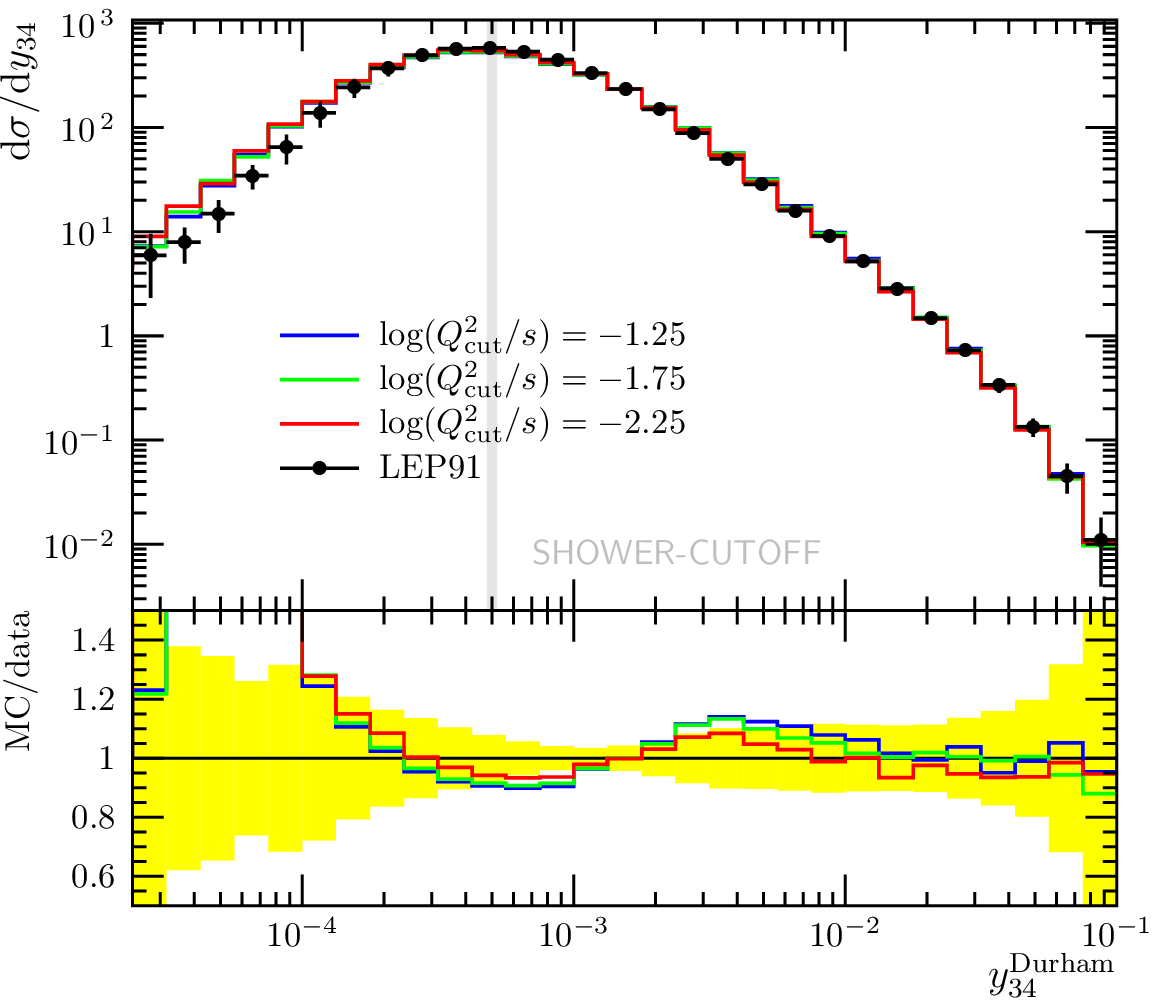}\vspace*{5mm}\\
  \includegraphics[width=0.485\linewidth]{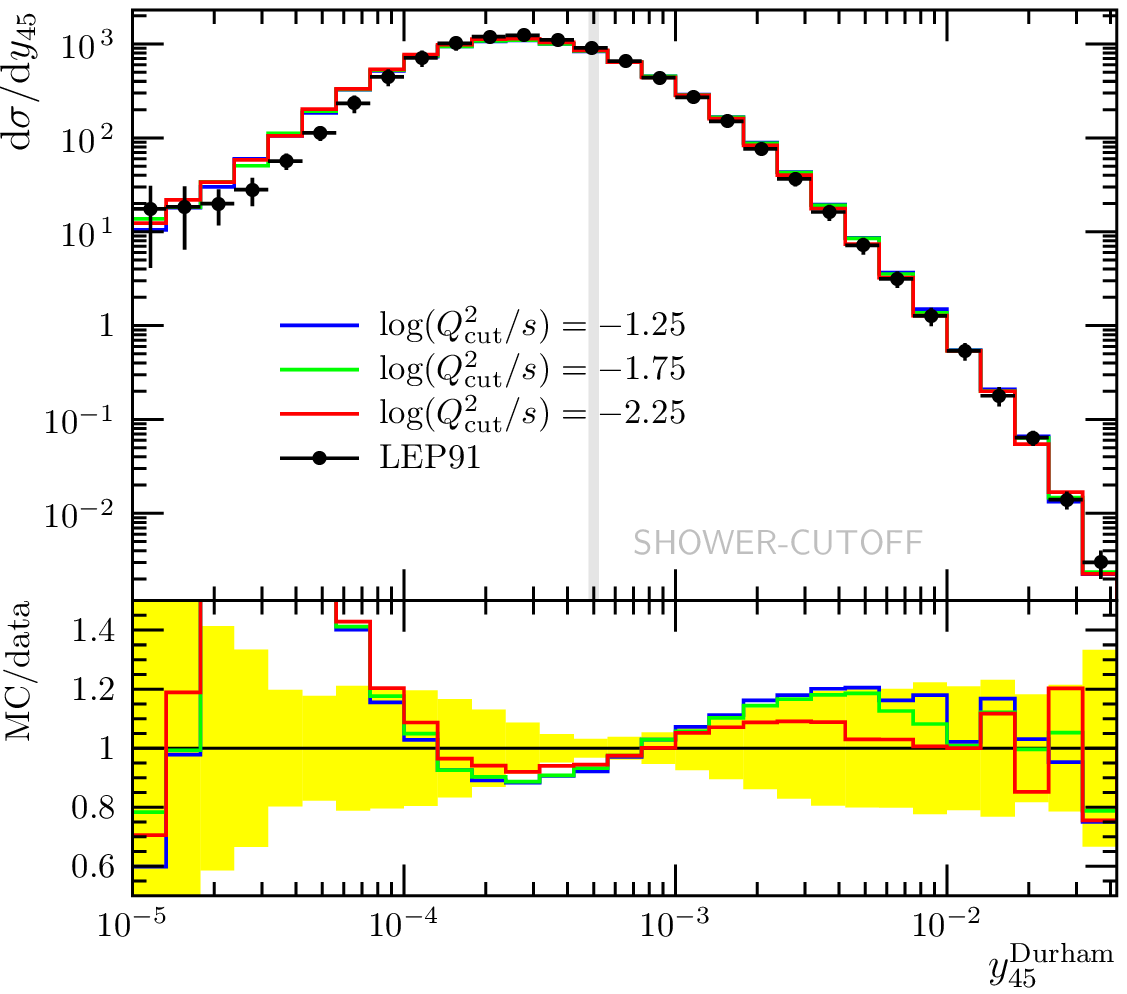}\hspace*{5mm}
  \includegraphics[width=0.485\linewidth]{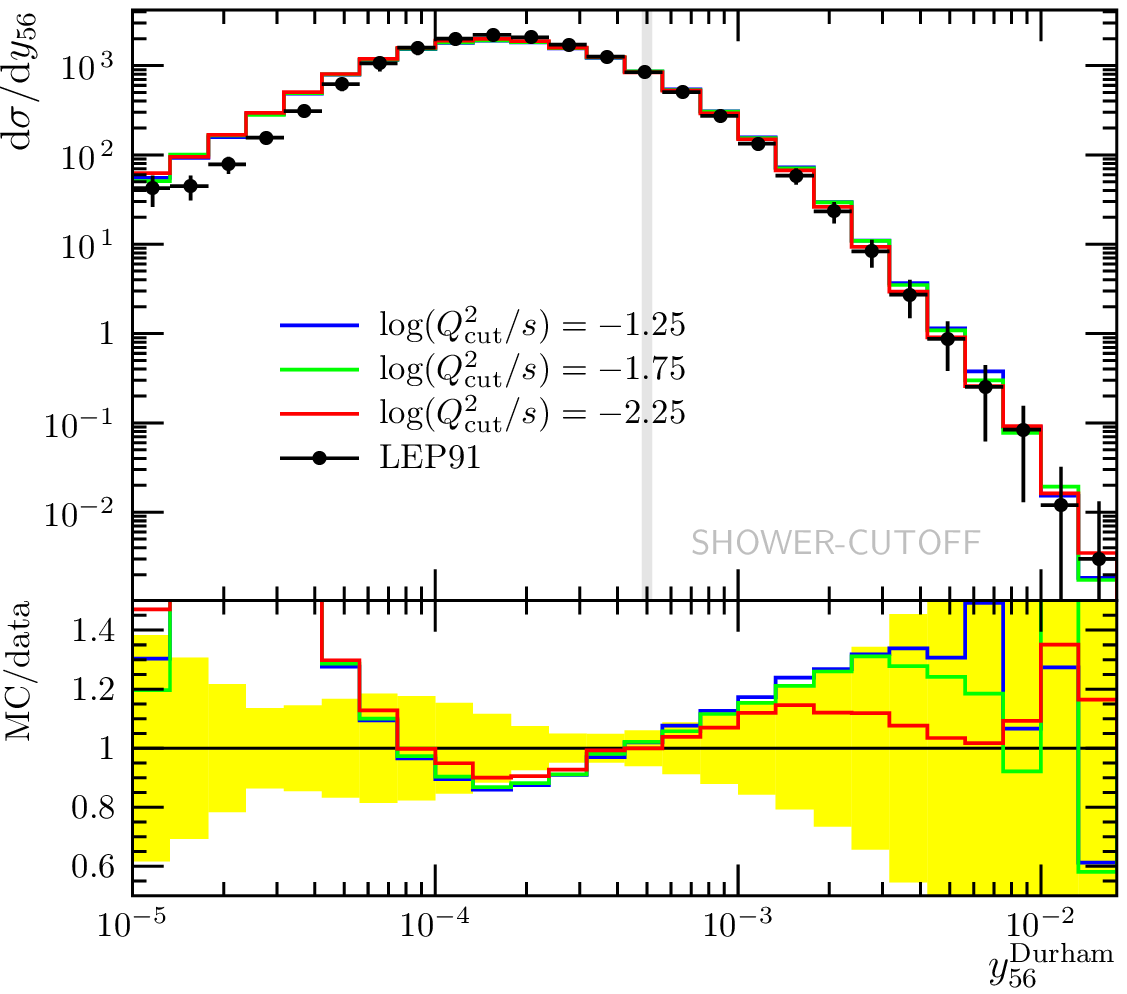}
  \caption{Differential jet rates for three different merging cuts compared
    to data from LEP\protect\cite{Pfeifenschneider:1999rz}.}
  \label{fig:eejetrates}
\end{centering}
\end{figure}

Figure~\ref{fig:eejetrates} shows the differential jet rates for a merged sample
of up to four additional jets from the matrix element, generated with \Comix 
and showered with the \CSS. The merging cuts, which have been used, are
$\logyc=-2.25$, $\logyc=-1.75$ and $\logyc=-1.25$. We observe only tiny deviations
between the predictions of the various samples.

\begin{figure}
\begin{centering}
  \includegraphics[width=0.315\linewidth]{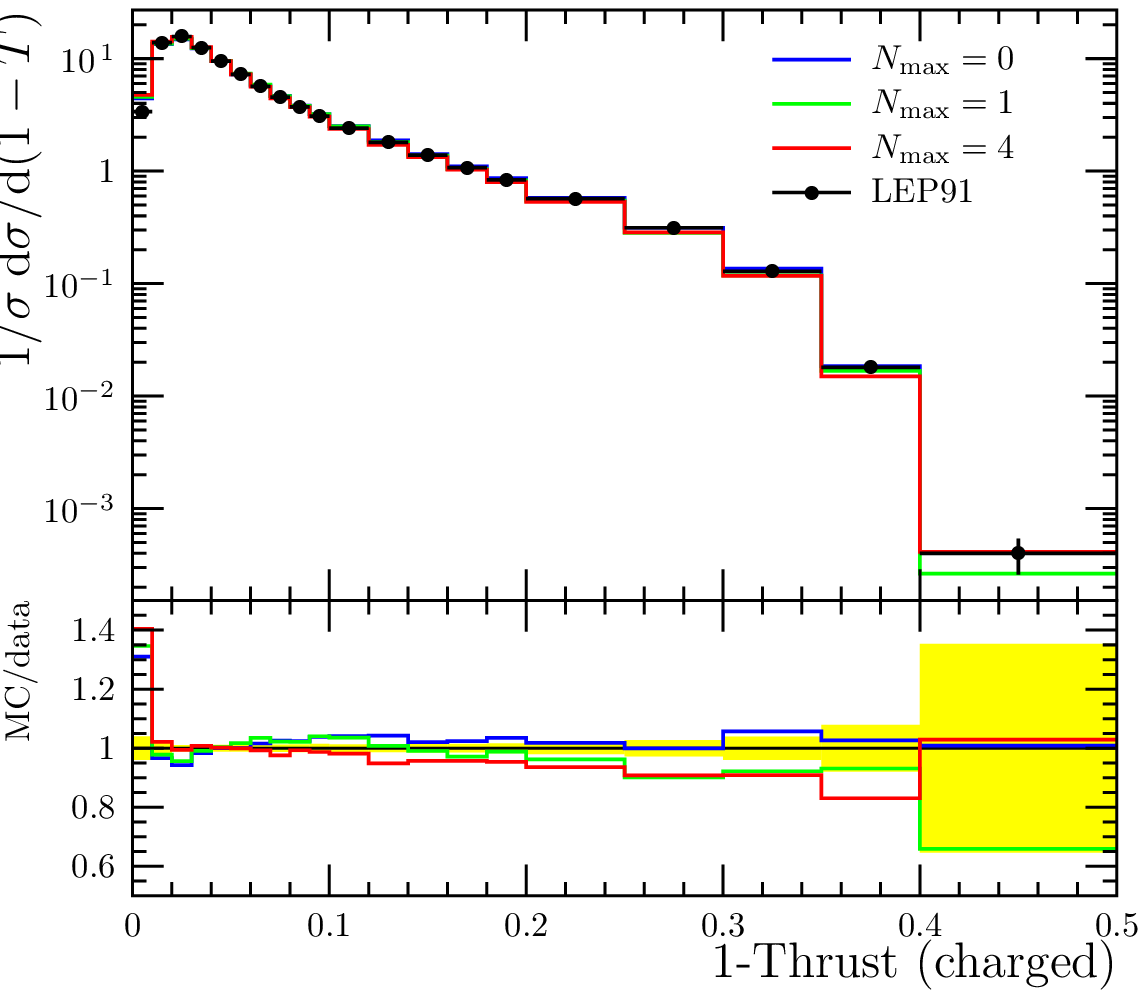}\hspace*{3mm}
  \includegraphics[width=0.315\linewidth]{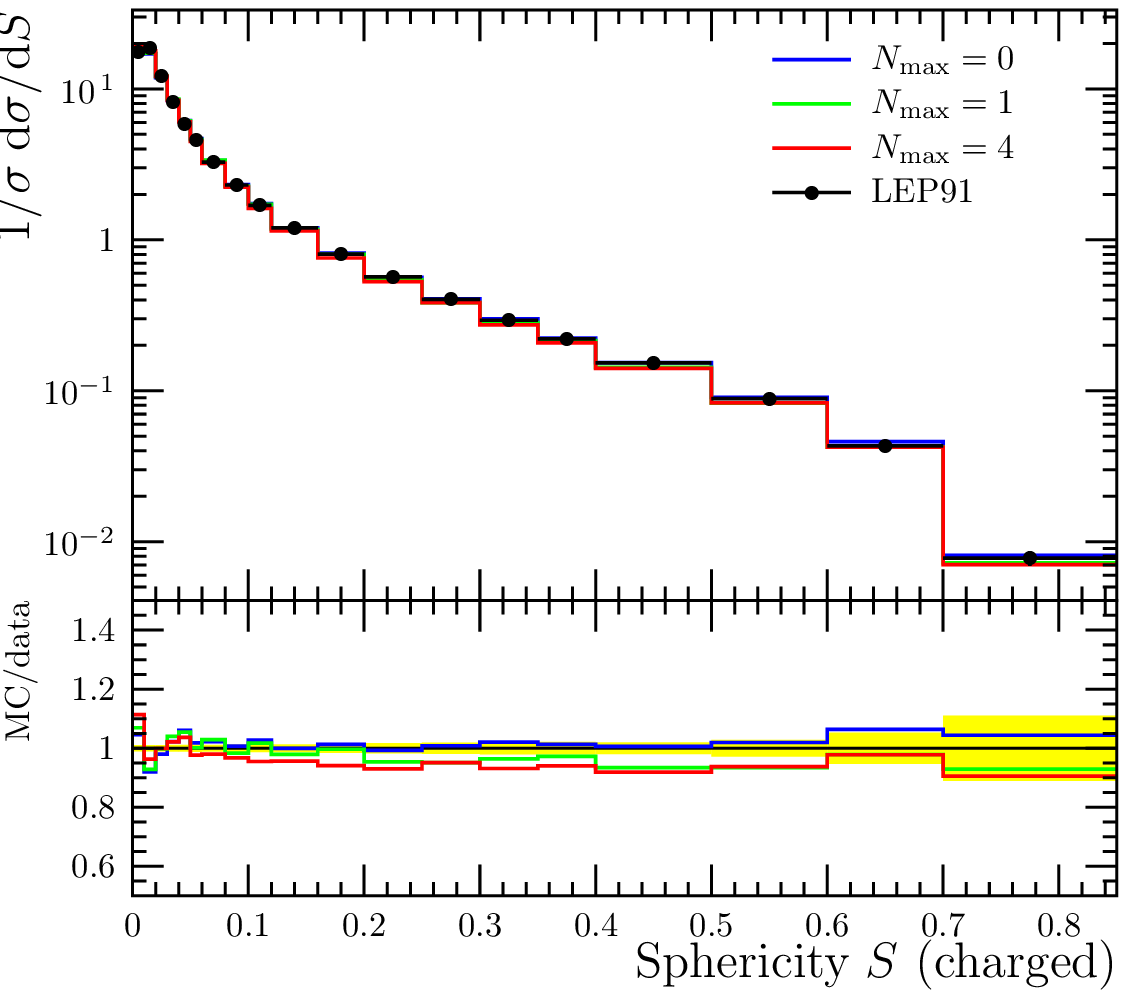}\hspace*{3mm}
  \includegraphics[width=0.315\linewidth]{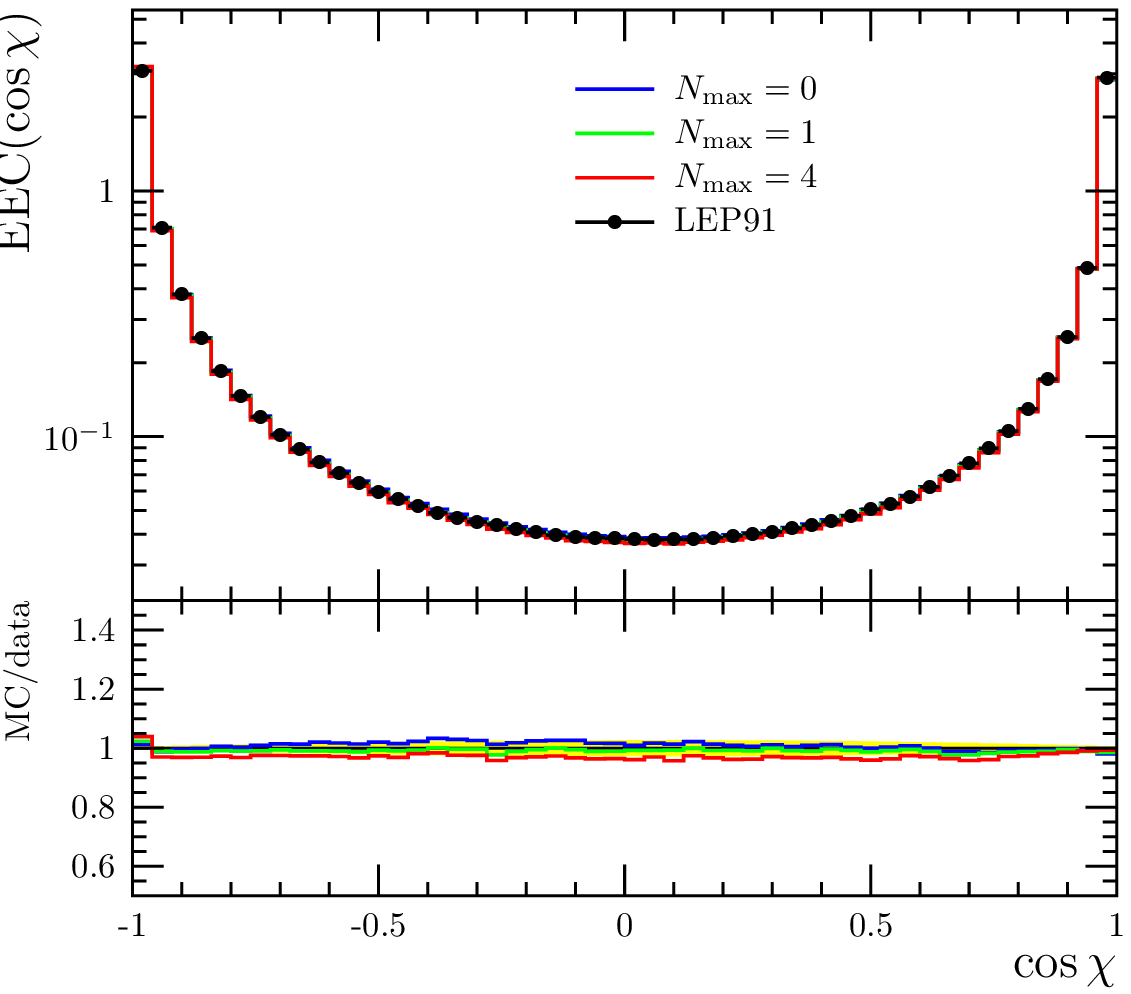}\hspace*{3mm}
  \caption{Event-shape observables compared to data from LEP\protect\cite{Abreu:1996na}.}
  \label{fig:eeeventshapes}
\end{centering}
\end{figure}

Event-shape observables like thrust, sphericity and the energy-energy
correlation are presented in Fig.~\ref{fig:eeeventshapes}. Details about their
definition and the corresponding data from experiment can be found
in~\cite{Abreu:1996na}.
These observables are well described by an appropriately tuned 
pure parton-shower setup already, and no matrix element improvement
is therefore necessary. On the other hand, the comparison between the pure
parton-shower sample and merged samples constitutes an important
consistency check.  We find very good agreement of the respective predictions.

\subsubsection*{Colour assignment at large $N_C$}
\begin{figure}
\begin{centering}
  \includegraphics[width=0.315\linewidth]{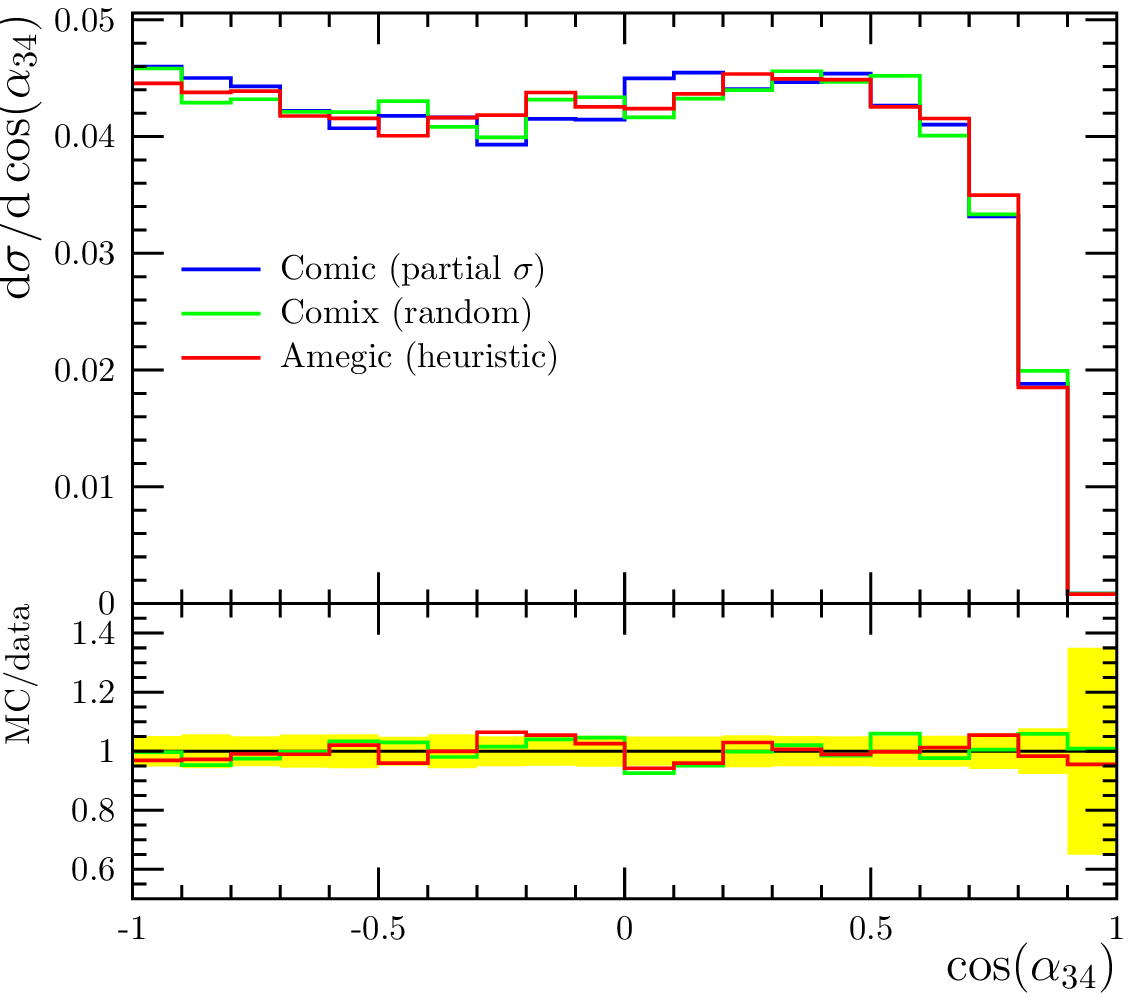}\hspace*{3mm}
  \includegraphics[width=0.325\linewidth]{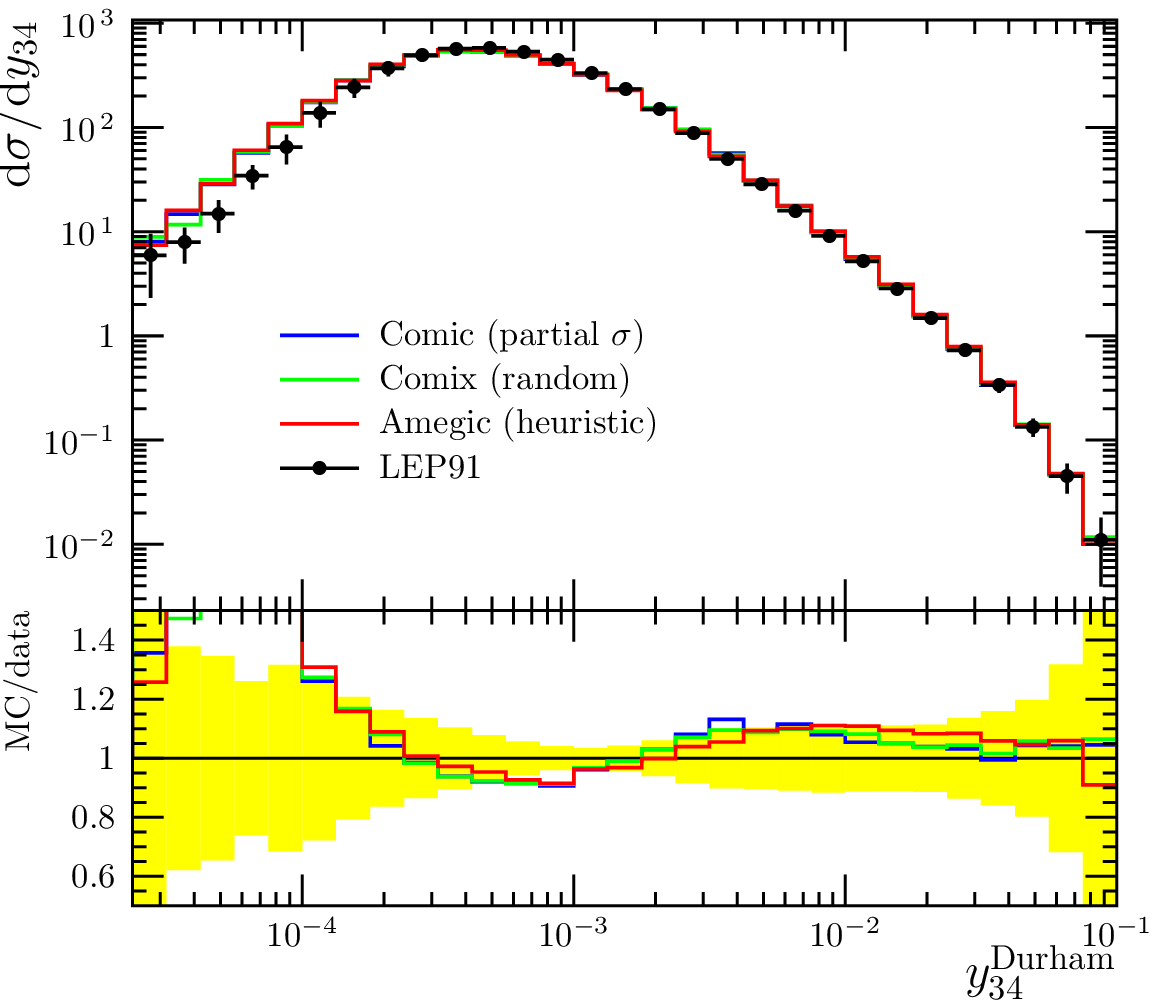}\hspace*{3mm}
  \includegraphics[width=0.315\linewidth]{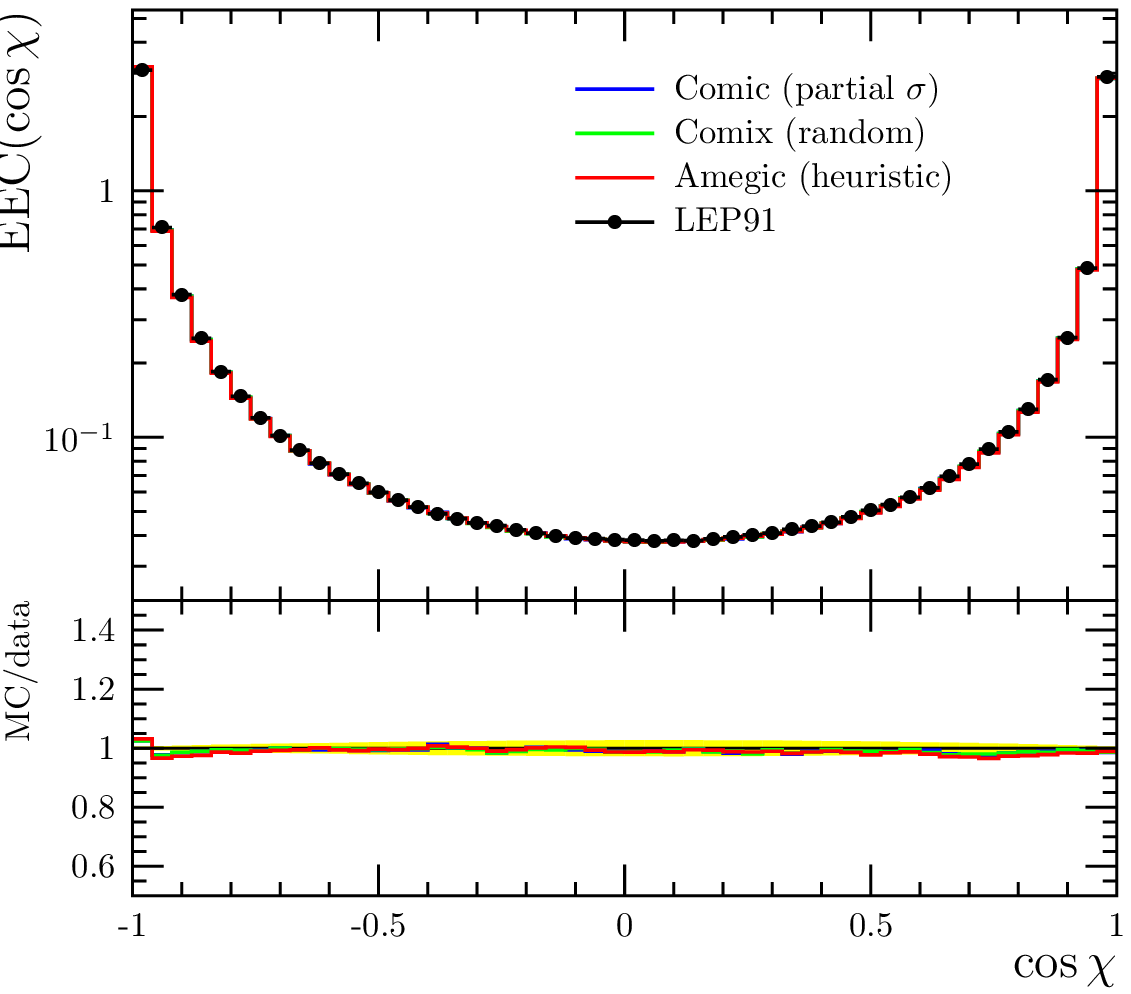}\hspace*{3mm}
  \caption{Different approaches to colour treatment and their effects 
    on $y_{34}$, $\alpha_{34}$ and EEC. Data are taken 
    from~\protect\cite{Pfeifenschneider:1999rz,Abreu:1996na}.}
  \label{fig:ee34}
\end{centering}
\end{figure}

To investigate the influence of different strategies to handle colour 
in matrix-element generation and merging, we compare the generators \Comix
and \Amegic. \Comix is run in two different modes, corresponding to the
two strategies of assigning colours in the large-$N_C$ limit explained in 
Sec.~\ref{sec:colour}. More precisely we denote by
\begin{description}
\item['partial $\sigma$'] \hfill \\
  selection of a colour assignment according to the
  proportionate squared partial amplitude corresponding to 
  this large-$N_C$ configuration (i.e.\ the {\em first} method
  presented in Sec.~\ref{sec:colour})
\item['random'] \hfill \\
  selection of a colour assignment according to the
  ratio of the squared partial amplitude corresponding to
  this large-$N_C$ configuration and the total amplitude squared
  (i.e.\ the {\em second} method presented in Sec.~\ref{sec:colour})
\item['heuristic'] \hfill \\
  heuristic colour assignment to the previously
  colour-summed amplitudes in \Amegic 
  (cf.\ Sec~\ref{sec:megenerators:amegic})
\end{description}
Figure~\ref{fig:ee34} shows respective predictions for the three
different choices.
Since the first configuration where different colour assignments
could take effect arises in four-jet events, the selected observables
are the $3 \to 4$ jet rate and the angle $\alpha_{34}$ between the two
softest jets, selected on an event-by-event basis.
Furthermore, the energy-energy correlation typically shows sensitivity to
the connection of the hadronisation phase to the parton-shower output, and could
thus depend on the colour setting as well.
We observe no significant differences between the various options.
This encourages to proceed with even the heuristic method, 
which enables us to employ the merging technique 
with various kinds of matrix element generators, including those
which do not allow a projection onto the large-$N_C$ limit.

\subsection{Drell-Yan lepton production in \texorpdfstring{$p\bar p$}{p pbar} collisions}
The scope of this section is to validate the proposed merging algorithm in 
collisions with hadronic initial states. One of the simplest processes
in this setup is Drell-Yan lepton pair production. It constitutes an
important testing ground to validate the applicability of the proposed
jet criterion and the interplay of the merging algorithm with the PDF's.

Event generation has been set up for $p\bar p$-collisions at a centre-of-mass energy of
$\sqrt{s}=1960 \,\GeV$. For the hard process a merged sample of 
$p\bar{p} \to e^+e^- + N\,{\rm jets}$ has been produced, where
$N \leq \Nmax$ with $0 \leq \Nmax \leq 6$. 
In addition, a cut of $66 \,\GeV < m_{e^+e^-} < 116 \,\GeV$ 
has been applied at the matrix-element level.
The factorisation scale has been chosen as $m_{e^+e^-}^2$. 
Note that the transverse mass squared of the lepton pair,
$m_{T,e^+e^-}^2=m_{e^+e^-}^2+{\rm k}_{\perp,e^+e^-}^2$, which is often selected 
as the factorisation scale in other merging approaches, is not a proper
choice for the proposed algorithm. It is non-continuous with respect to
the transverse momentum, ${\rm k}_{\perp,e^+e^-}^2$, because
the leading-order configuration is generated with ${\rm k}_{\perp,e^+e^-}^2=0$
and the minimum transverse momentum of events with one additional jet
is limited by the phase-space separation cut.

\subsubsection*{Total cross sections and jet rates}
Again, we first present a comparison of total cross sections predicted by the merging algorithm
for various values of the separation criterion $\qcut$ and the maximum jet multiplicity $\Nmax$.
Table~\ref{tab:zjetsxs} shows the respective results.
Differences range up to 2.3\%, between the leading-order cross section and predictions for the 
merged samples.
Usually, the systematic uncertainties in hadronic collisions are larger than in
$e^+e^-$ reactions, partly due to PDF uncertainties and partly because the shower evolution 
generated by the \CSS and the resummation used to compute the PDF's are not entirely compatible, 
cf.~\cite{Catani:1996vz}. This effect could account for slightly larger deviations between 
results for the merged samples. However, we observe a reasonably low variation.
\mytable{tbp}{
\begin{tabular}{cc|c|r@{}l|r@{}l|r@{}l|r@{}l|r@{}l|r@{}l|}
\cline{3-15}
                    & & \multicolumn{13}{c|}{$\Nmax$} \\ \cline{3-15}
                    &    & 0              & \mct{1}   & \mct{2}   & \mct{3}   & \mct{4}    & \mct{5}    & \mct{6}   \\ \hline
\mco{\mrt{$\qcut$}} & 20 GeV & \mrt{192.6(1)} & 191&.0(3) & 190&.5(4) & 189&.0(5) & 189&.4(7)  & 188&.2(8) & 189&.9(10) \\ \cline{2-2}\cline{4-15}
\mco{}              & 30 GeV &                & 192&.3(2) & 192&.7(2) & 192&.6(3) & 192&.9(3)  & 192&.7(3)  & 193&.2(3) \\ \cline{2-2}\cline{4-15}
\mco{}              & 45 GeV &                & 193&.6(1) & 194&.4(1) & 194&.3(1) & 194&.4(1)  & 194&.6(2)  & 194&.4(1) \\ \hline
\end{tabular}
}{Total cross sections [pb] in $p \bar{p} \to e^+ e^- + \rm{jets}$ at
  $\sqrt{s}=1960 \,\GeV$ and their dependence on the merging cut.
  \label{tab:zjetsxs}}

\begin{figure}
\begin{centering}
  \includegraphics[width=0.7\linewidth]{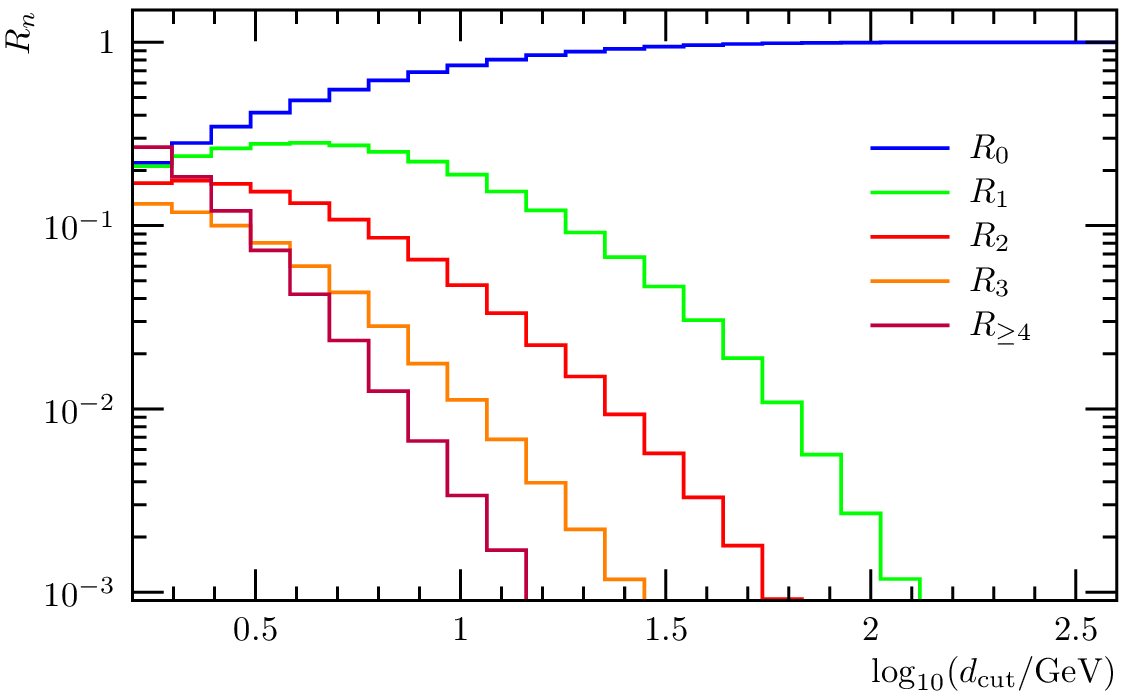}\\
  \caption{Integrated jet rates as a function of the analysis cut.}
  \label{fig:zjetsnjetrates}
\end{centering}
\end{figure}

Figure~\ref{fig:zjetsnjetrates} shows integrated rates of jets determined 
with the CDF Run~II $k_T$-algorithm~\cite{Blazey:2000qt} as a function of
the analysis cut $d_{\rm cut}$.
Monte Carlo results have been produced using a merged sample for up to five jets 
in the final state, generated with \Comix and showered with the \CSS with a
merging cut $\qcut=20 \GeV$.

\subsubsection*{Differential distributions}
To study the merging systematics in more detail, we investigate again
the differential jet rates $d_{n\,n+1}$, describing the scales at which 
$n+1$ jets are clustered into $n$ jets according to the CDF Run~II 
$k_T$-algorithm. This algorithm has a free parameter, $D$, which
accounts for the missing information on beam partons. 
Hence, in this setup, no firm relation can be established between the jet measure
of the $k_T$-algorithm and the jet criterion, Eq.~\eqref{eq:jet_criterion}.
Nevertheless, a certain correspondence between the two quantities exists,
making these distributions a good testing ground for variations around the
merging cuts.

\begin{figure}
\begin{centering}
  \includegraphics[width=0.485\linewidth]{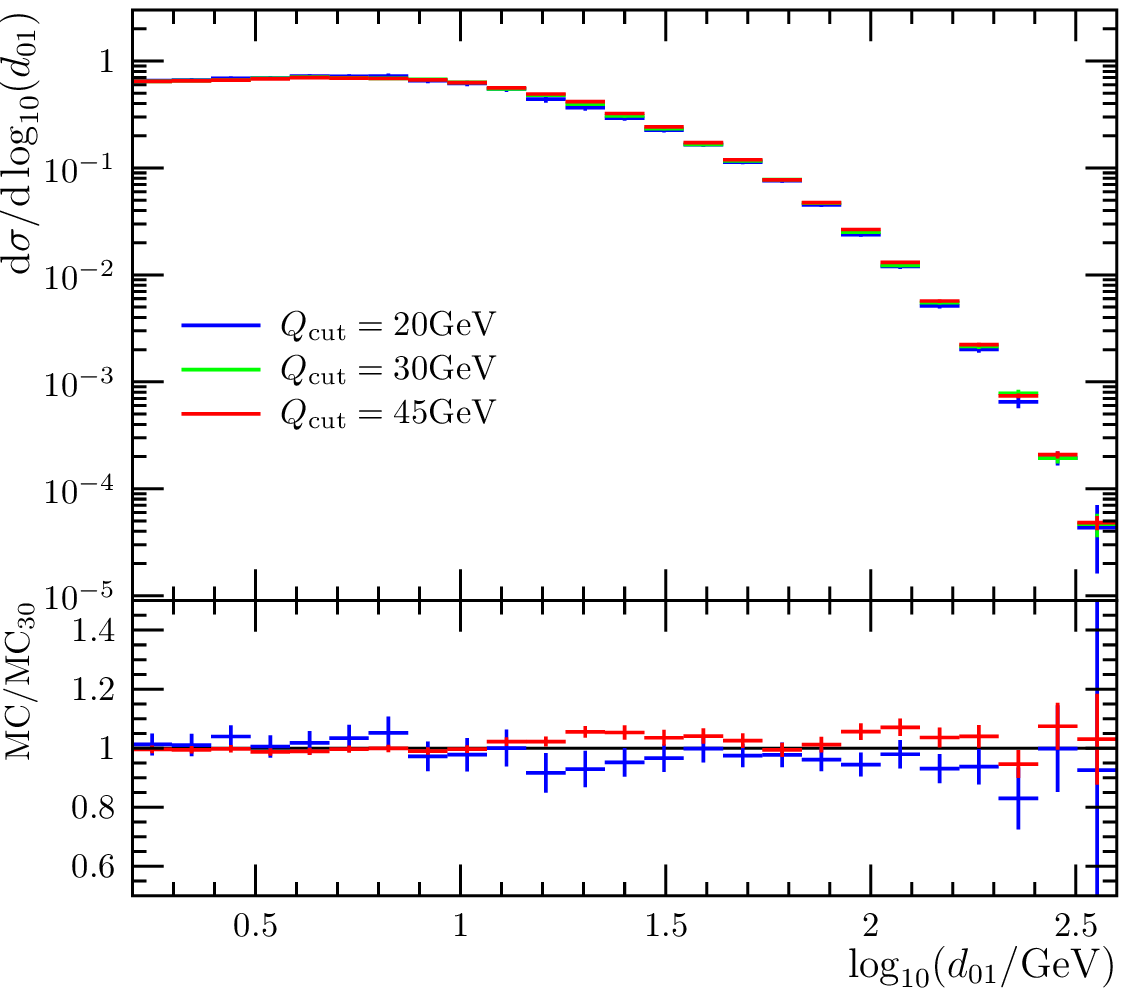}\hspace*{5mm}
  \includegraphics[width=0.485\linewidth]{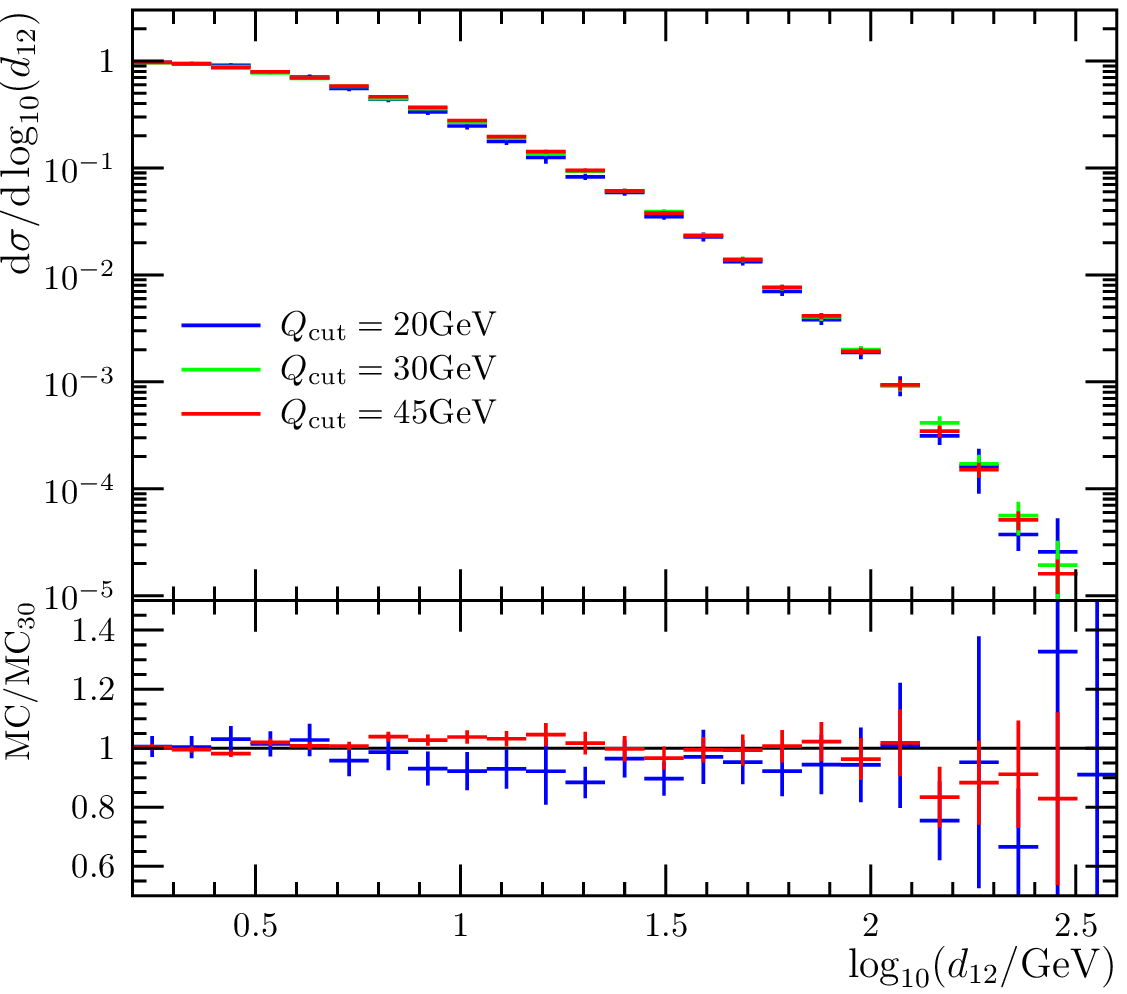}\vspace*{5mm}\\
  \includegraphics[width=0.485\linewidth]{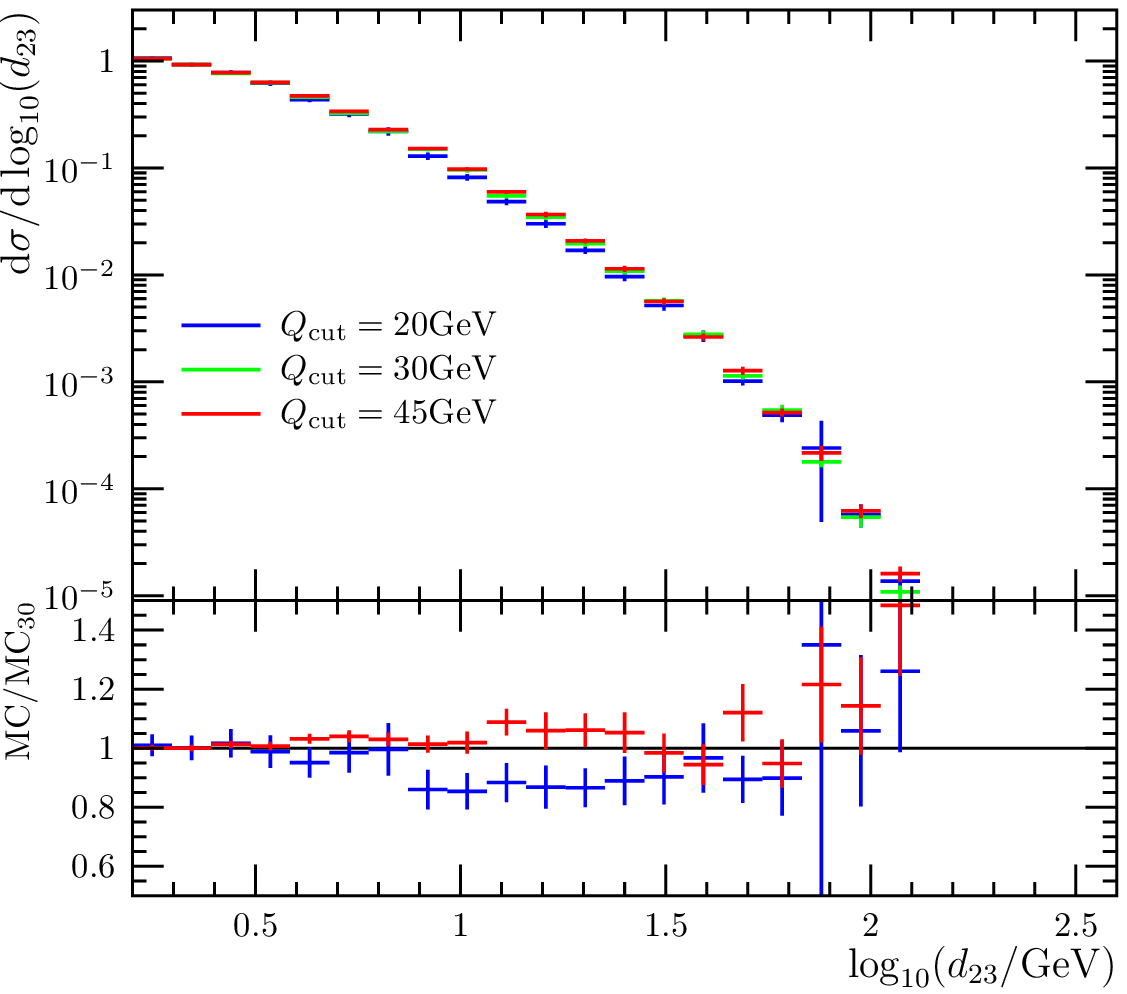}\hspace*{5mm}
  \includegraphics[width=0.485\linewidth]{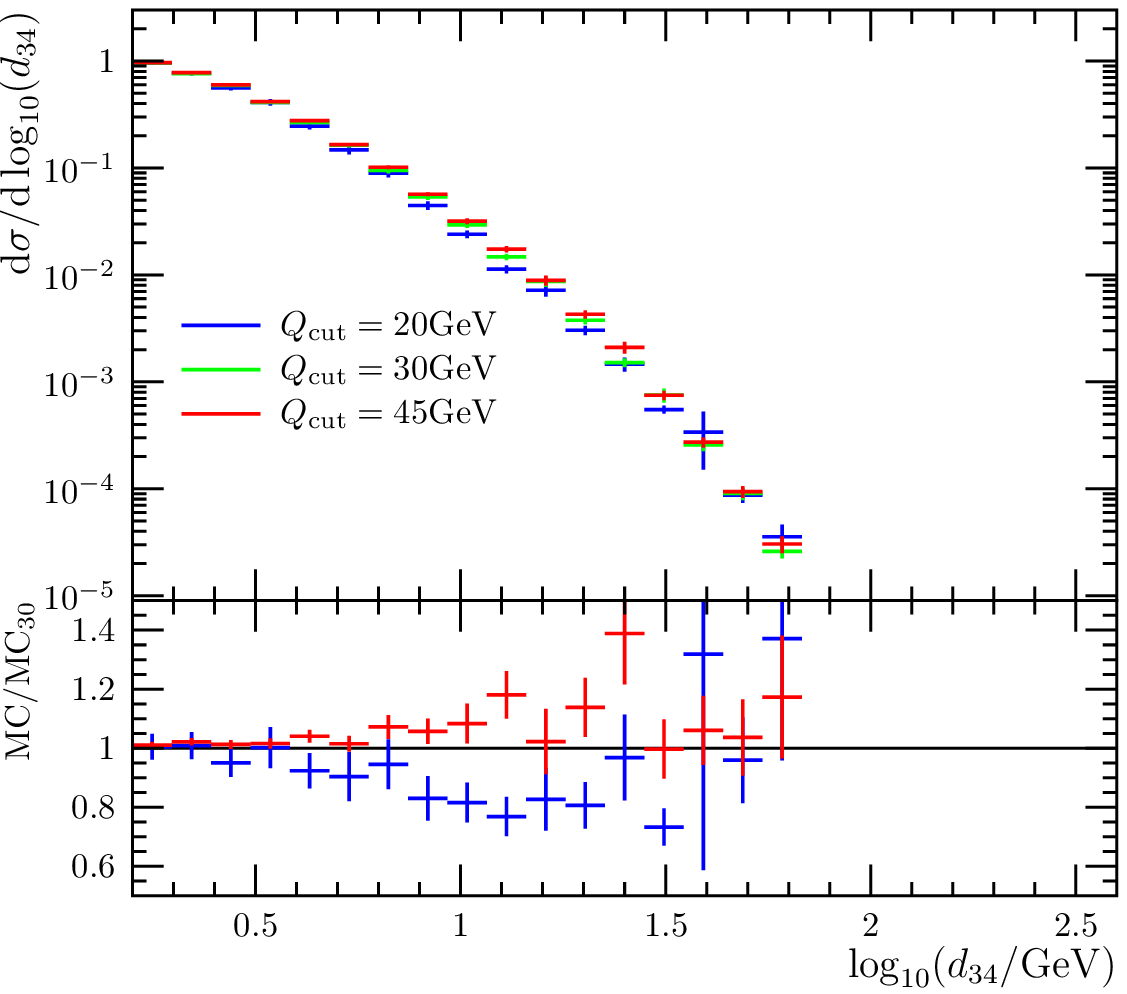}\\
  \caption{Differential jet rates $d_{n\,n+1}$ for three different merging cuts.}
  \label{fig:zjetsjetrates}
\end{centering}
\end{figure}

To produce Fig.~\ref{fig:zjetsjetrates} a merged sample
of up to five jets from the matrix element has been generated with \Comix 
and showered with the \CSS. The merging cuts, which have been used, are
$\qcut=20 \,\GeV$, $\qcut=30 \,\GeV$, and $\qcut=45 \,\GeV$.
As in the case of $e^+e^-$ collisions, the deviations 
between the predictions of the various samples are small.

\begin{figure}
\begin{centering}
  \includegraphics[width=0.485\linewidth]{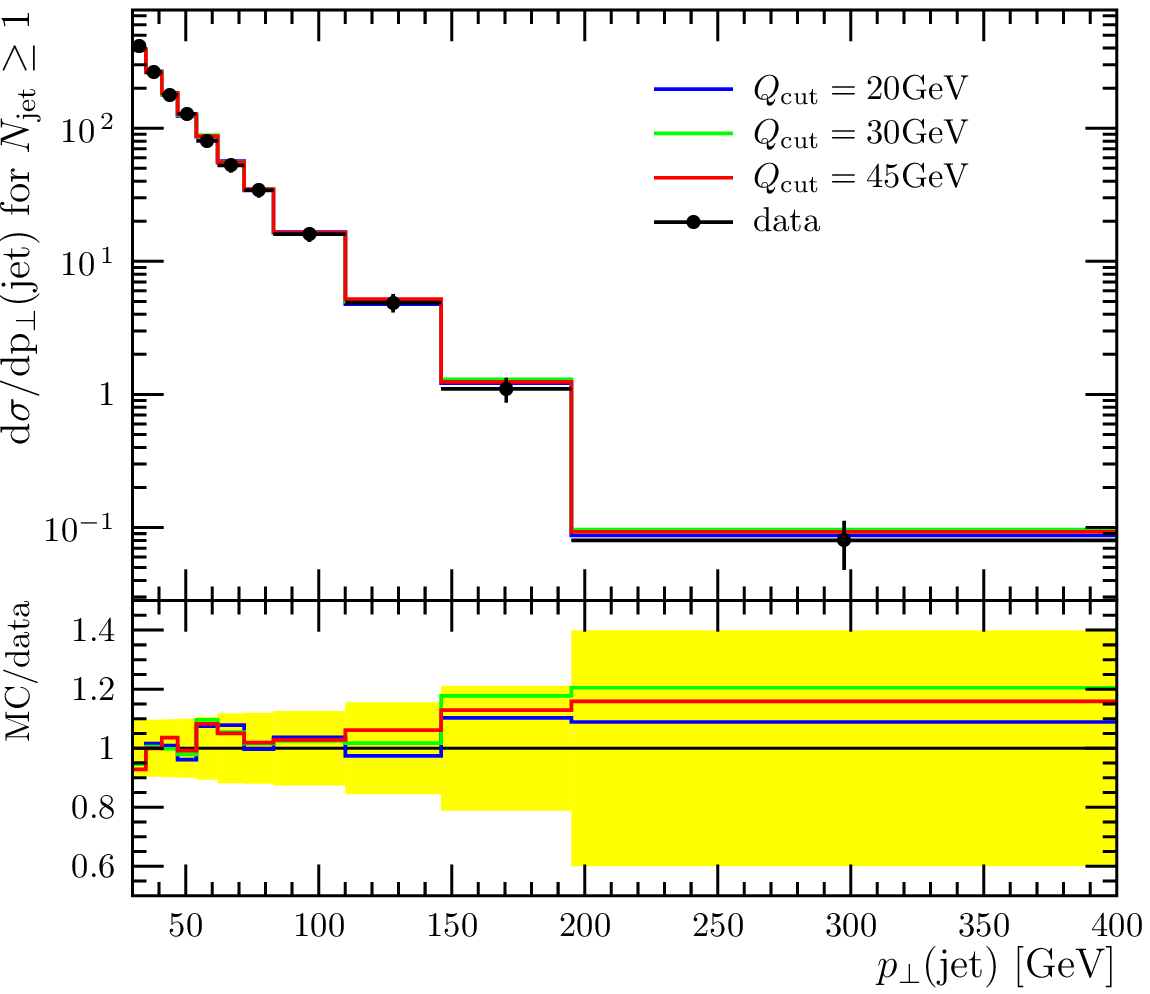}\hspace*{5mm}
  \includegraphics[width=0.485\linewidth]{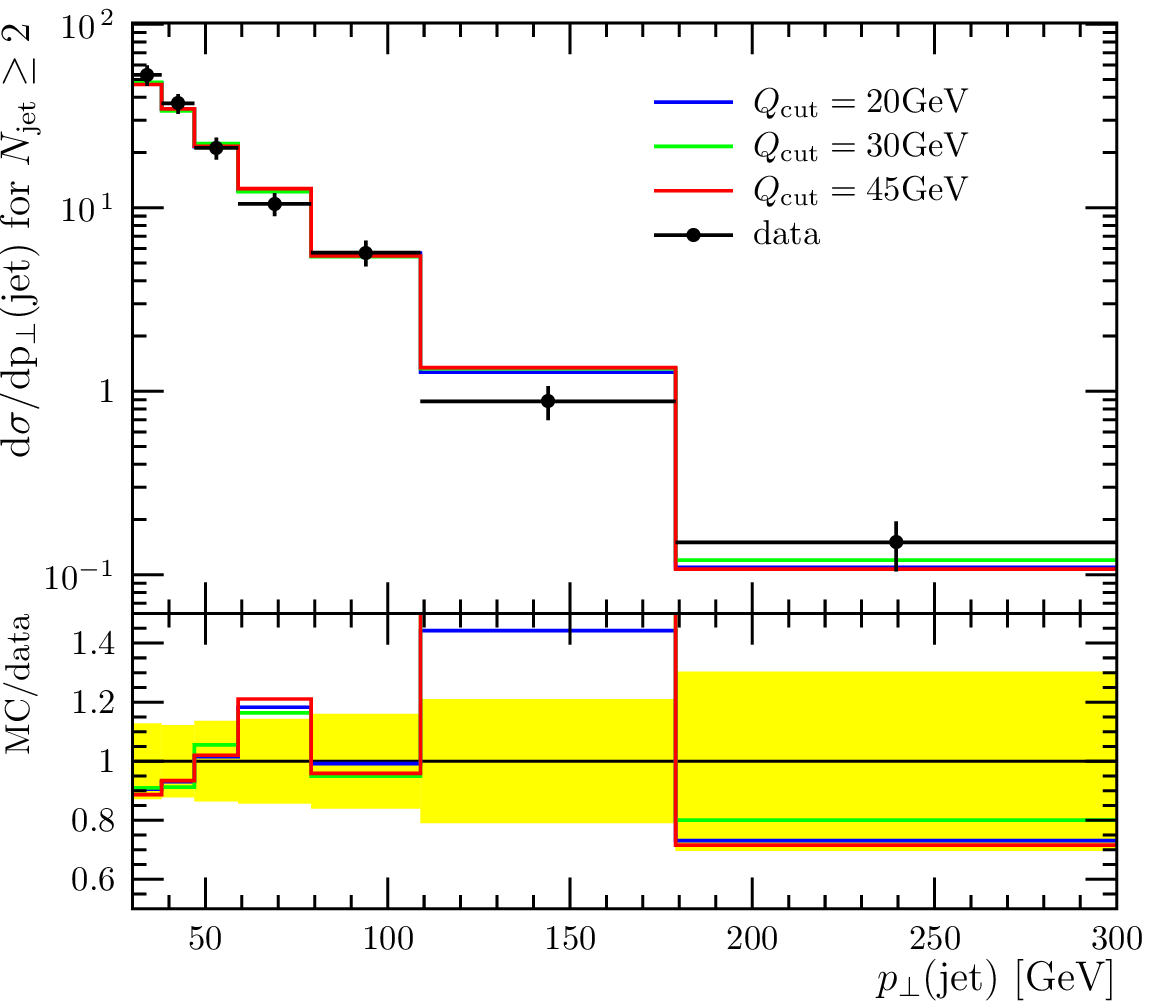}\\
  \caption{Jet $p_T$ in $N_{\rm jet} \ge 1$ and $N_{\rm jet} \ge 2$ events compared to data from CDF\protect\cite{:2007cp}.}
  \label{fig:zjetsjetpts}
\end{centering}
\end{figure}

Most observables are even less sensitive to the 
precise value of the merging cut. As an example,
Figure~\ref{fig:zjetsjetpts} displays the transverse momentum of the two leading jets for
the three merging cut values in comparison to data from
CDF\cite{:2007cp}.

It is also interesting to understand the influence of the maximal number of
jets generated from the matrix element, $\Nmax$, on experimental observables.
We observe that typically the predictions are fairly stable for the
$\Nmax$ leading jets. To put it another way, for a given analysis
investigating the $n$'th jet, one should use a Monte Carlo sample with
$\Nmax \geq n$. Due to the increased phase space available for QCD radiation
at the LHC, the higher jet multiplicities will play an even more important
role there.

\begin{figure}
\begin{centering}
  \includegraphics[width=0.485\linewidth]{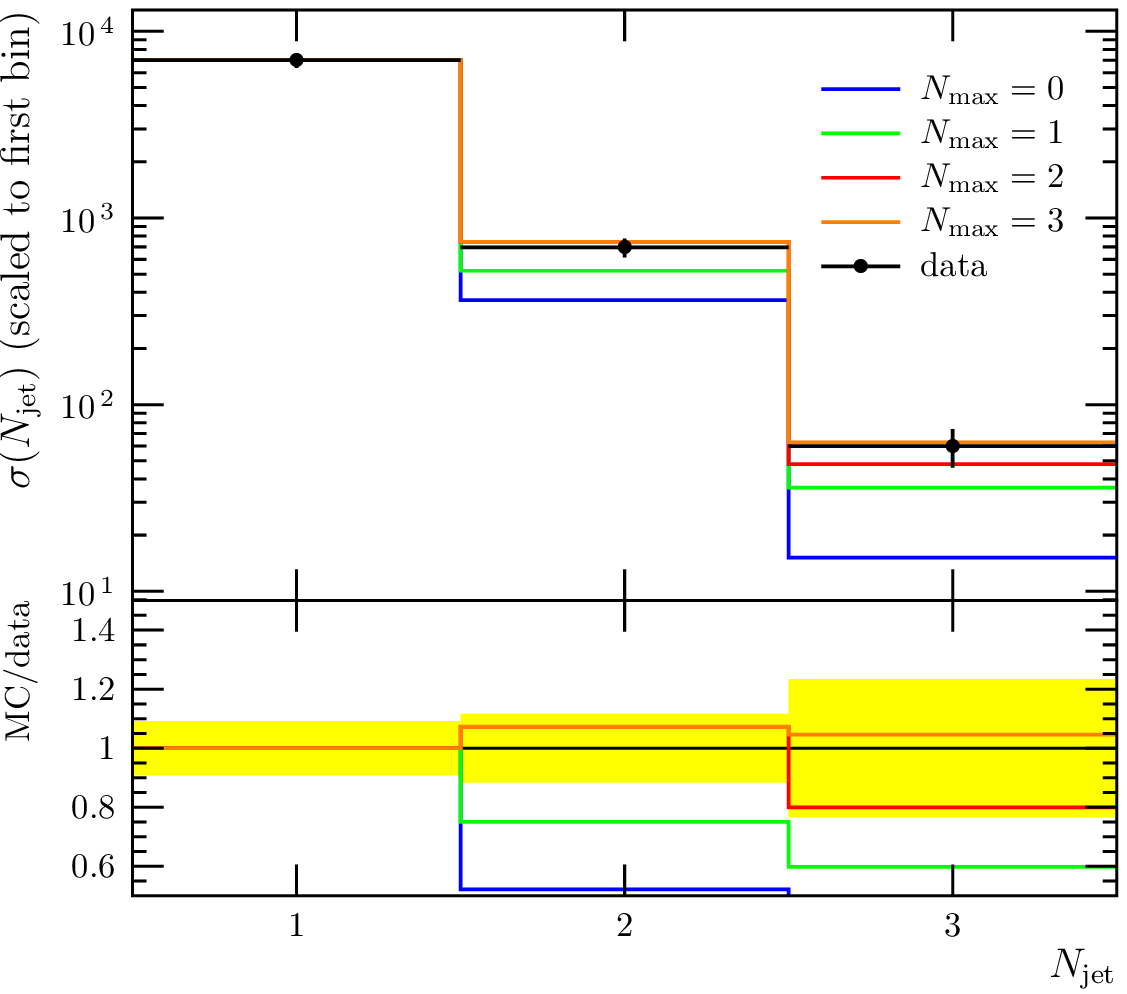}\hspace*{5mm}
  \includegraphics[width=0.485\linewidth]{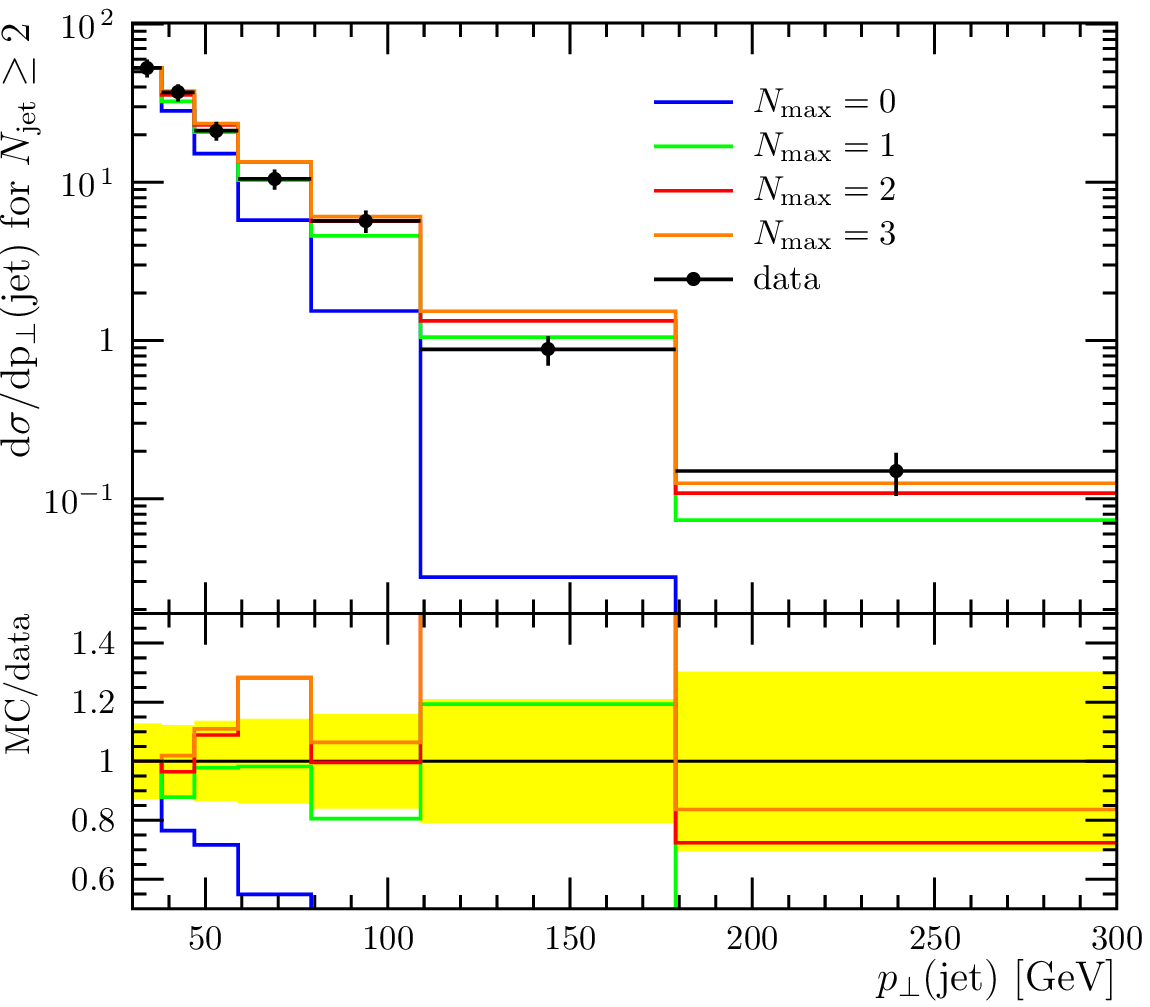}\\
  \caption{Jet multiplicity and jet $p_T$ in $N_{\rm jet} \ge 2$ events compared to data from CDF\protect\cite{:2007cp}.}
  \label{fig:zjetscdfnjet}
\end{centering}
\end{figure}

Again, comparing to data from CDF\cite{:2007cp} in Figure~\ref{fig:zjetscdfnjet}
and varying $\Nmax$ between zero and three, the importance of correctly describing
additional hard jet production by the respective matrix elements can be estimated.

\section{Conclusions}
\label{sec:conclusions}

In this publication we have presented a general formal framework to discuss algorithms 
for the merging of multi-jet matrix elements and parton showers.  We have
constructed a merging algorithm that maintains the
logarithmic accuracy provided by the parton shower in both initial and final state radiation.  
In this construction,
special emphasis is put on an invariant formulation of the respective phase-space 
separation criterion.  Because this criterion is not identical with the parton-shower evolution
parameter, the logarithmic accuracy can only be maintained by running a truncated shower. 

Hard matrix elements must be interpreted in the large-$N_C$ limit to provide an input 
for shower Monte Carlos. Since the respective strategy is not unambiguous, the influence 
of different methods to assign colours was studied.  We find no significant difference 
between the proposed algorithms, which range from heuristic assignment to the choice 
of a configuration with probability proportional to the respective colour ordered 
subamplitude squared.

We have checked the systematics of the newly proposed algorithm in $e^+e^-$ annihilation
into hadrons and in Drell-Yan lepton pair production.  We find greatly reduced 
uncertainties, compared to previously employed methods like the original CKKW technique.  
This statement holds for inclusive quantities such as total cross sections and jet rates, 
as well as for differential distributions.

\section*{Acknowledgements}
We would like to thank Jan Winter, Tanju Gleisberg and Marek Sch{\"o}nherr for
collaboration on the \Sherpa project, which provided the framework for our study.
Special thanks go to them as well as Thomas Binoth and Nigel Glover for fruitful discussions
and their comments on the manuscript. We are grateful to Hendrik Hoeth and Andy Buckley 
for their help with \Rivet,
which we used for the analyses and the comparison with data.  SH acknowledges 
funding by the HEPTOOLS Marie Curie Research Training Network (contract number 
MRTN-CT-2006-035505) and the Swiss National Science Foundation (SNF, contract 
number 200020-117602).  FS gratefully acknowledges financial support by the MCnet 
Marie Curie Research Training Network (contract number MRTN-CT-2006-035606).
The work of SS was partially funded by the STFC.

\appendix
\section{Relation to other merging methods}
\label{sec:relation}

In this appendix we discuss, on formal grounds, the relation of the merging 
algorithm proposed here with the original CKKW method~\cite{Catani:2001cc}.  
Its relation with the CKKW-L method, presented in~\cite{Lonnblad:2001iq,*Lavesson:2005xu}, 
and with the MLM method~\cite{Mangano:2001xp} will also briefly be discussed,
arguing on more algorithmic grounds.

\subsection{Formal equivalence with the CKKW method at NLL accuracy}
\label{sec:ckkw}
Let us start by showing the formal equivalence, at next-to-leading logarithmic
accuracy, of the CKKW method presented in~\cite{Catani:2001cc} and the proposed
merging algorithm. For this purpose, the generating function $\phi_a\rbr{Q,q\,;\phi^{(0)}}$ 
for a jet at scale $q$, produced by a jet that emerged at scale $Q$ with initial condition
$\phi^{(0)}$ is computed using the coherent-branching formalism. 
We focus on massless partons. All jet scales are given in terms of the Durham jet measure, 
which is defined as~\cite{Brown:1991hx,*Catani:1991hj}
\begin{equation}\label{eq:jet_measure_durham}
  Q_{ij}^2=2\min\cbr{E_i^2,E_j^2}\,\rbr{1-\cos\theta_{ij}}
    =2\,p_ip_j\,\min\cbr{\frac{\tilde{z}_{i,j}}{1-\tilde{z}_{i,j}},
      \frac{1-\tilde{z}_{i,j}}{\tilde{z}_{i,j}}}\;,
  \quad\text{where}\quad \tilde{z}_{i,j}=\frac{E_i}{E_i+E_j}\;.
\end{equation}
Note that in the collinear region and for massless particles, the jet criterion,
Eq.~\eqref{eq:jet_criterion}, effectively encodes the Durham measure for all splittings
which are singular in $\tilde{z}_{i,j}$ or $1-\tilde{z}_{i,j}$.
It is therefore justified to employ the Durham measure instead throughout this computation.

Collinear factorisation is assumed while the evolution parameter
is the angular variable
\begin{equation}\label{eq:angular_evolution_variable}
  t\to\tilde{q}^{\,2}=\frac{k_\perp^2}{z^2(1-z)^2}=\frac{p^2}{z(1-z)}\;,
\end{equation}
with $z$ the light-cone momentum fraction in the splitting
and $p^2$ the virtuality of the mother parton. 

Comparing the coherent-branching formalism to an arbitrary shower algorithm
respecting colour coherence, potential differences arise because coherence
is implemented through angular ordering. However, with regard to QCD evolution
in the collinear regime, both methods are formally equivalent.

The generating function is defined by the differential equation
\begin{equation}\label{eq:jet_generating_function_diff}
  \begin{split}
  \frac{\done\phi_a\rbr{q,Q_0;\phi^{(0)}}}{\done\log(q/Q_0)}\,=&\;\int_0^1\done z\,
    \sum\limits_{b=q,g}\mc{K}_{ab}(z,q)\,\Theta\sbr{\,\min\cbr{z,1-z}q-Q_0\,}\\
  &\quad\times\sbr{\vphantom{\sum}\,\phi_b\rbr{zq,Q_0;\phi^{(0)}}\,
    \phi_c\rbr{(1-z)q,Q_0;\phi^{(0)}}-\phi_a\rbr{q,Q_0;\phi^{(0)}}\,}\;.
  \end{split}
\end{equation}
where 
\begin{equation}
  \mc{K}_{ab}(z,q)\,=\;\frac{\alpha_s(z(1-z)\,q)}{2\pi}\,P_{a\to bc}(z)\;.
\end{equation}
The upper bounds $zq$ and $(1-z)q$ in $\phi_b$ and $\phi_c$ account for 
the modified starting conditions in the evolution of the respective
jet~\cite{Catani:1990rr}. The functions $\phi^{(0)}$ represent the initial condition
for the evolution.

Equation~\eqref{eq:jet_generating_function_diff} can be cast into a more
suitable form by employing Sudakov form factors in terms of the Durham
jet measure, computed at next-to-leading logarithmic accuracy,
cf.~\cite{Brown:1991hx,*Catani:1991hj}. For the purpose of this analysis 
we include the finite terms, such that
\begin{equation}\label{eq:nll_sudakov}
  \begin{split}
  \bar\Delta_q(q,Q)=&\;\exp\cbr{-\int_q^Q\frac{\done\bar q}{\bar q}\,
    \int_0^{1-q/Q}\done z\;2\,\mc{K}_{qq}(z,q)\,}\\
  \bar\Delta_g(q,Q)=&\;\exp\cbr{-\int_q^Q\frac{\done\bar q}{\bar q}\,
    \sbr{\;\int_{q/Q}^{1-q/Q}\done z\,\mc{K}_{gg}(z,q)+
      N_f\int_0^1\done z\,\mc{K}_{gq}(z,q)\,}\,}\;.
  \end{split}
\end{equation}

Thus,
\begin{equation}\label{eq:jet_generating_function_sud}
  \begin{split}
  &\frac{\done}{\done\log(q/Q_0)}\,\frac{\phi_a\rbr{q,Q_0;\phi^{(0)}}}{\bar\Delta_a(Q_0,q)}\,=\;
  \frac{1}{\bar\Delta_a(Q_0,q)}\int_0^1\done z\,\sum\limits_{b=q,g}\mc{K}_{ab}(z,q)\\
  &\text{\hspace*{10ex}}\times\Theta\sbr{\,\min
      \cbr{z,1-z}\,q-Q_0\,}\,\phi_b\rbr{zq,Q_0;\phi^{(0)}}\,
    \phi_c\rbr{(1-z)q,Q_0;\phi^{(0)}}\;.
  \end{split}  
\end{equation}
For quark jets, employing that the largest contribution to the branching $q\to q$ 
comes from the region $1-z\ll 1$,
\begin{equation}\label{eq:jet_generating_function_quark_diff}
  \frac{\done}{\done\log(q/Q_0)}\,
  \log\frac{\phi_q\rbr{q,Q_0;\phi^{(0)}}}{\bar\Delta_q(Q_0,q)}\,\approx\;
  \int_0^{1-Q_0/q}\done z\;2\,\mc{K}_{qq}(z,q)\,\phi_g\rbr{(1-z)q,Q_0;\phi^{(0)}}\;.
\end{equation}
Note that the lower limit on the $z$ integration is redefined to equal zero,
because the largest contribution to the integral arises from the region
$z\approx 1$. Using $\phi_q\rbr{Q_0,Q_0;\phi^{(0)}}=\phi_q^{(0)}$ and
inserting the quark Sudakov form factor, yields
\begin{equation}\label{eq:jet_generating_function_quark}
  \phi_q\rbr{Q,Q_0;\phi^{(0)}}\,\approx\;\phi_q^{(0)}\,\exp\cbr{\,
  \int_{Q_0}^Q\done q\,\int_0^{1-q/Q}\done z\;2\,\mc{K}_{qq}(z,q)\,
    \sbr{\,\phi_g\rbr{q,Q_0;\phi^{(0)}}-1\,}\,}\;.
\end{equation}
The solution to the respective evolution equation for gluon jets
can be found in~\cite{Brown:1991hx,*Catani:1991hj}.

As discussed in~\cite{Catani:2001cc}, a vetoed parton shower can be expressed 
in terms of a different starting condition, 
$\phi^{(0)}\to \tilde{\phi}(Q,Q_{\rm cut},Q_0;\phi^{(0)})$.
Then, in order to generate the correct jet fractions at some scale $Q_0$, the
following identity is imposed:
\begin{equation}
  \phi_a(Q,Q_{\rm cut};\tilde{\phi})\,\overset{!}{=}\;
  \phi_a\rbr{Q,Q_0;\phi^{(0)}}\;.
\end{equation}
Substituting the generating function $\phi_q$, 
Eq.~\eqref{eq:jet_generating_function_quark}, yields, for the quark
generating function,
\begin{equation}\label{eq:genfunc_vetoedps_plus_cutme}
  \begin{split}
  &\phi_q\rbr{Q,Q_0;\phi^{(0)}}\,=\;\tilde{\phi}_q(Q,Q_{\rm cut},Q_0;\phi^{(0)})\\
  &\text{\hspace*{10ex}}\times\exp\cbr{\,
  \int_{Q_{\rm cut}}^Q\done q\,\int_0^{1-q/Q}\done z\;2\,\mc{K}_{qq}(z,q)\,
  \sbr{\,\phi_g\rbr{q,Q_0;\phi^{(0)}}-1\,}\,}\;.
  \end{split}
\end{equation}
Consequently, the result for the modified parton shower reads
\begin{equation}\label{eq:genfunc_vetoedps_result}
  \begin{split}
  &\tilde{\phi}_q(Q,Q_{\rm cut},Q_0;\phi^{(0)})\,=\;\phi^{(0)}
  \exp\cbr{\,\int_{Q_0}^{Q_{\rm cut}}\done q\,\int_0^{1-q/Q}\done z\;2\,\mc{K}_{qq}(z,q)\,
  \sbr{\,\phi_g\rbr{q,Q_0;\phi^{(0)}}-1\,}\,}\;.
  \end{split}
\end{equation}
At next-to-leading logarithmic accuracy this is equivalent to the
phase-space slicing method introduced in Sec.~\ref{sec:merging}, because
Eqs.~\eqref{eq:genfunc_vetoedps_plus_cutme} and~\eqref{eq:genfunc_vetoedps_result}
can be written as
\begin{equation}\label{eq:genfunc_vetoedps_plus_cutme_new}
  \begin{split}
  &\phi_q\rbr{Q,Q_0;\phi^{(0)}}\,=\;\tilde{\phi}_q(Q,Q_{\rm cut},Q_0;\phi^{(0)})\\
  &\text{\hspace*{10ex}}\times\exp\cbr{\,
  \int_{Q_0}^Q\done q\,\int_0^{1-q/Q}\done z\;2\,\mc{K}^{\rm ME}_{qq}(z,q)\,
  \sbr{\,\phi_g\rbr{q,Q_0;\phi^{(0)}}-1\,}\,}
  \end{split}
\end{equation}
and
\begin{equation}\label{eq:genfunc_vetoedps_result_new}
  \begin{split}
  &\tilde{\phi}_q(Q,Q_{\rm cut},Q_0;\phi^{(0)})\,=\;\phi^{(0)}
  \exp\cbr{\,\int_{Q_0}^Q\done q\int_0^{1-q/Q}\done z\;2\,\mc{K}^{\rm PS}_{qq}(z,q)\,
  \sbr{\,\phi_g\rbr{q,Q_0;\phi^{(0)}}-1\,}\,}\;.
  \end{split}
\end{equation}
Here, the combination of an extended integration range for $q$ and the cutoff 
due to the $\Theta$-functions contained in $\mc{K}^{\rm ME}$ and $\mc{K}^{\rm PS}$ 
generates exactly the same $q$-integral as in 
Eqs.~\eqref{eq:genfunc_vetoedps_plus_cutme} and~\eqref{eq:genfunc_vetoedps_result}.
A similar proof holds for the gluon-jet generating function.

Note that the extended range for $z$-integration in 
Eq.~\eqref{eq:jet_generating_function_quark_diff} would have been generated in a 
natural way by the jet criterion, Eq.~\eqref{eq:jet_criterion}. For $q\to qg$
splittings the $\Theta$-function corresponding to the one in 
Eq.~\eqref{eq:jet_generating_function_diff} would read $\Theta\sbr{(1-z)q-Q_0}$,
thus setting the integration boundaries $0\le z\le 1-q/Q_0$. In any case the
contribution from the range $0\le z\le q/Q_0$ is subleading.

\subsubsection*{Algorithmic differences}

Arguing on more algorithmic grounds, the following differences between
the CKKW method and the new merging prescription presented in this publication 
can be established:
\begin{itemize}
\item Usually, Sudakov form factors within the CKKW approach are implemented 
  in the form of (semi-)analytic functions, see for example~\cite{Krauss:2004bs}.
  The accuracy of the method is then limited by the level of correspondence 
  between these functions and the respective parton-shower result.
\item Truncated showers, which, according to Sec.~\ref{sec:truncated_showers},
  are mandatory to establish formal equivalence of the evolution in the pure parton shower
  and the matrix-element improved Monte Carlo algorithm, are implemented in an approximate way.
  They are replaced by modified starting conditions for external partons, rather
  than emissions from internal lines. The net effect is, that (up to typically 
  less important momentum reshuffling) no redefinition of matrix element kinematics is necessary.
\end{itemize}

\subsection{Correspondence to the CKKW-L method}

In order to see the relation of the algorithm proposed here with the CKKW-L 
prescription presented in~\cite{Lonnblad:2001iq,*Lavesson:2005xu}, let us
concentrate first on the case of pure final-state radiation.  A prime example
for this is $e^+e^-$-annihilations into hadrons.  In this case, CKKW-L, based on 
\Ariadne~\cite{Lonnblad:1992tz}, the jet criterion  and the (dipole) shower-evolution 
parameter are identical.  In fact, jets are defined by the dipole transverse momentum 
$p_\perp$ also used in the shower.  For a splitting $\tilde i\tilde k\to ijk$, 
\begin{equation}
  p_\perp^2 = \frac{s_{ij}s_{jk}}{s_{\tilde i\tilde k}}\,.
\end{equation}
Because both quantities coincide, there is no need for a truncated shower: Whenever
an emission is harder than that of a node given by the hard matrix element,
i.e.\ at a larger $p_\perp$, then it will produce a new jet.  Since such emissions
lead to vetoing the event, no intermediate emissions, which are not giving rise to 
a jet, must be taken into account.  In this case the algorithm for $e^+e^-$ 
annihilations reduces to
\begin{enumerate}
\item[$\bullet$] Relevant multi-jet cross sections for the process under consideration
  are calculated with the phase-space restriction $Q>Q_{\rm cut}$. Strong couplings 
  are computed such that they give an overestimate, which can be reweighted. 
\item Events are generated according to the above defined 
  cross sections with kinematics determined by the respective matrix elements.
\item The most probable shower history of the final state is determined through backward 
  clustering, cf.\ Sec.~\ref{sec:cluster_algo}. The clustering is guided by information 
  from the matrix element, which means that only those shower histories 
  may be identified, which have a corresponding Feynman diagram.
\item
  The event is accepted or rejected according to a kinematics-dependent 
  weight, which corresponds to evaluating strong couplings in the shower scheme
  and computing the no-branching probability for each dipole.  This is done by
  starting the shower from the scale at which the dipole emerged.  If the first 
  emission generated by the shower is harder than the scale given by the next
  matrix-element generated node related to that dipole (dipole contains at least one 
  internal leg), or harder than $Q_{\rm cut}$ (dipole made of two external legs),
  then the event is rejected.  
\end{enumerate}
It is apparent that this is in perfect agreement with the method proposed in this
paper, if the additional complication of the truncated shower can be neglected, which 
becomes obsolete if the shower-evolution parameter and the jet criterion coincide.

Comparing our method with the CKKW-L method in case of hadron collisions is not
that simple.  In principle, the extension of the algorithm for the dipole
shower presented above to the case of hadron collisions is straightforward, 
especially with identical jet criterion and shower-evolution parameter.  However,
due to the fact that \Ariadne re-interprets initial-state radiation as final-state
radiation with one or both dipole legs being the proton remnant, it is not 
entirely clear, how PDF effects are accounted for.  An obvious way out would be
to replace this non-perturbative algorithm for initial-state radiation with a
formulation that rests entirely on the grounds of perturbation theory and the
factorisation theorems, like the one presented in~\cite{Winter:2007ye}.  This
opens an arena for further interesting studies.

\subsection{Relation with the MLM method}

It is difficult to establish any formal correspondence between the method proposed 
here and the MLM prescription~\cite{Mangano:2001xp}.  This is mainly due to the fact that 
they indeed base on different ideas and methods.  To our understanding, though, the
main difference lies in the treatment of radiation off intermediate legs.  Therefore
we will very briefly describe the MLM method here and we will outline the apparent
differences with respect to the new approach.

The MLM prescription, also presented in~\cite{Hoche:2006ph,*Alwall:2007fs} bases on using a 
simple cone definition as the jet criterion to generate the parton configurations at 
the matrix-element level.  The scales of strong couplings are reconstructed by 
using a backward clustering with a $k_\perp$ measure.  Then the accepted configurations 
are passed on to parton-shower routines, typically the ones of \Pythia~\cite{Sjostrand:2006za} 
(virtuality or $p_\perp$-ordered shower) or of \HERWIG~\cite{Corcella:2002jc}
(angular-ordered shower).  These codes typically reconstruct the parton-shower 
starting scales of a multi-parton configuration by directly inspecting colour 
connections, without any backward clustering and therefore partially neglecting 
intermediate legs and the radiation originating from them.  Having the starting
conditions at hand, the shower is invoked without any constraint.  After it has 
terminated at the hadronisation scale the original partons stemming from the matrix 
element are matched to the jets present at parton-shower level, again defined by the 
cone algorithm.  If such a match is not possible, either due to extra, unwanted jets 
being produced in the parton shower or due to ``loosing'' jets in the parton shower, 
the event is rejected.  

The differences to the other algorithms are obvious: 
\begin{itemize}
\item While the MLM prescription computes the Sudakov suppression weight in an 
  inclusive way, the other algorithms, the original CKKW approach, the CKKW-L method 
  and the algorithm proposed here, determine the Sudakov rejection by combining the 
  rejection weights from the individual partons.  Therefore, the effect of
  ``loosing'' jets is not present, and only the emergence of unwanted jets
  leads to vetoing the event.  
\item Because the definition of starting scales for the parton shower is left to 
  the routines provided by \Pythia and \HERWIG it is not entirely clear, how far
  radiation of intermediate legs is accounted for.  

  In any case, however, at present, neither of the two codes, \Pythia and \HERWIG
  allows truncated showering.  This, following our argument, would be mandatory to
  formally guarantee the logarithmic accuracy provided by the respective parton-shower
  algorithm.
\end{itemize}
Despite these differences, a good agreement of predictions obtained with the CKKW 
method and the MLM method can often be observed. Respective results have been reported 
for example in~\cite{Hoche:2006ph,*Alwall:2007fs}.
They indicate, in particular, that the differences between the two methods may not be 
too important in the case of $W$+jets production at the Tevatron and the LHC.
It can, however, not be expected that this statement holds for arbitrary processes.

\section{Merging with multiple leading-order processes}
\label{sec:multicore_merging}
In this appendix, we consider situations like Drell-Yan production 
at large centre-of-mass energies.
In such configurations the scale of the leading-order process -- in our case the lepton 
pair production process -- is low compared to the potential scales of additional
jets. Hence, simply applying the pure merging prescription introduced in
Sec.~\ref{sec:merging} will eventually result in missing hard radiation and
therefore not lead to a good description of experimentally relevant signatures.
An example for such signatures would be a lepton pair with mass lower than
the transverse momentum of an accompanying QCD jet.
\begin{stress}
Note, that this is consistent with the proposed merging approach, 
since the choice of a specific factorisation scale at the leading order
corresponds to generating additional QCD radiation up to an evolution parameter 
limited by this scale. Instead, the solution lies in the dynamic definition 
of a leading-order process for each event.
\end{stress}
Let us stick to the example of Drell-Yan production. 
Shower branchings cannot take place at scales larger than the 
factorisation scale, in this case usually the mass of the Drell-Yan pair, $m_{ll}$. 
Jets harder than this scale should thus actually be described by 
a leading-order process corresponding to ``Drell-Yan + jet''-production. 
However, one needs to preserve the integrity of the inclusive Drell-Yan sample 
up to its factorisation scale. 

The event-generation algorithm allows to do so. The central idea is to 
reinterpret the leading-order process as soon as the scale of jet production 
in any matrix-element configuration with additional jets exceeds the 
factorisation scale. The corresponding event is then not
taken into account for the inclusive Drell-Yan production sample,
but is internally treated as part of the ``Drell-Yan + hard jet''-regime.
Its factorisation scale and the corresponding shower starting conditions 
are redefined accordingly.

Note that the problem discussed in this appendix only occurs for leading-order 
configurations which do not include any strong coupling\footnote{
  The case of effective theories, where QCD partons couple to non-QCD particles 
  via loop graphs, which are in turn reinterpreted as effective vertices,
  is not considered here.}.
Also, jets at scales lower than the mass of the Drell-Yan pair are always
well described in the merging. Such configurations can be produced by the parton shower,
in principle at arbitrary multiplicity, because of the highest multiplicity
treatment introduced in Sec.~\ref{sec:highest_multi}.

The relevance of the problem and the difference between the pure merging approach 
and the refined approach with multiple leading-order processes depends on the scales 
which are involved. If the phase space available to additional parton emission
is not too large, like in the case of $W$ or $Z$ production at the Tevatron,
resummation effects are small, because the difference between factorisation 
scale and the scale of hard extra jet production is usually small. In this case, 
the pure merging algorithm can safely be employed, since not too many hard jets 
can be generated, and this limited hard extra-jet multiplicity can be accounted 
for by matrix elements.

\bibliographystyle{bib/amsunsrt_mod}  
\bibliography{bib/journal}
\end{document}